\documentclass[12pt]{iopart}
\usepackage{iopams}

\usepackage{graphics,epsfig}
\usepackage{subfigure}
\usepackage{color,psfrag}
\usepackage{pgf,tikz}

\newcommand{\bin}[2]{C^{#2}_{#1}}
\newcommand{\nn}{\nonumber}
\newcommand{\bb}[0]{\begin{eqnarray}}
\newcommand{\ee}[0]{\end{eqnarray}}

\newcommand{\efig}[1]{Fig.~\ref{#1}}

\newcommand{\ff}{\frac{1}{2}}

\newcommand{\Sp}[2]{\left [#1\right ]_{#2}}
\newcommand{\sga}{\sigma_1^{\alpha}}
\newcommand{\sgb}{\sigma_2^{\alpha}}
\newcommand{\ac}{K'}
\newcommand{\acd}{\gamma}
\newcommand{\he}{\frac{h_e}{T}}

\newcommand{\free}[1]{F^{(#1)}}
\newcommand{\en}[1]{E^{(#1)}}
\newcommand{\ent}[1]{S^{(#1)}}
\newcommand{\cv}[1]{C_v^{(#1)}}
\newcommand{\suscept}[1]{\chi^{(#1)}}

\newcommand{\Kr}[2]{\tilde K_{#2}^{(#1)}}
\newcommand{\Ir}[2]{\tilde I_{#2}^{(#1)}}
\newcommand{\Hr}[2]{\tilde H_{#2}^{(#1)}}
\newcommand{\tw}{\widetilde w}
\newcommand{\DF}{\Delta_F}

\begin{document}
\title{Random site dilution properties of frustrated magnets on a hierarchical lattice}

\author{Jean-Yves Fortin}
\address{CNRS, Institut Jean Lamour, D\'epartement de Physique de la Mati\`ere
et des
Mat\'eriaux,
UMR 7198, Vandoeuvre-les-Nancy, F-54506, France\\
}
\date{\today}
\ead{jean-yves.fortin@univ-lorraine.fr}
\begin{abstract}
We present a method to analyze magnetic properties of frustrated Ising spin models
on specific hierarchical lattices with random dilution. Disorder is induced by dilution and geometrical
frustration rather than randomness in the internal couplings of the original
Hamiltonian. The two-dimensional model presented here possesses a macroscopic
entropy at zero temperature in the large size limit, very close
to the Pauling estimate for spin-ice on pyrochlore lattice, and a crossover
towards a paramagnetic phase. The disorder due to dilution is
taken into account by considering a replicated version of the recursion
equations between partition functions at different lattice sizes.
An analysis at first order in replica number allows for a systematic
reorganization of the disorder configurations, leading to a recurrence scheme.
This method is numerically implemented to evaluate the thermodynamical 
quantities such as specific heat and susceptibility in an
external field.
\end{abstract}

%\keywords{Grassmann algebra | partition function | dimer}
%\pacs{05.20.Gg,05.50.+q,75.40.Cx}

\maketitle

\section{Introduction}

Hierarchical lattices possess some interesting features of geometrical scale invariance, and several methods
have been developed  extensively in the past to study thermodynamical properties
of spin models considered on such structures~\cite{kaufman81,griffiths82}. The
similarity with Migdal-Kadanoff renormalization procedure for magnetic spin
systems~\cite{berker79} with exact recursion relations between coupling
constants at the
successive stages of lattice construction
makes such models suitable for the study of critical properties. Indeed, phase
transitions in these systems have non-mean field critical exponents, effective
dimensions and frustration effects within local loops. However, the absence of
translation and the site dependent connectivity inherent to such theoretical
construction make the physics apparently different from real systems. The
importance of exact solutions in disordered lattices are highlighted by the
possibility to obtain controllable recursion relations which gives interesting
properties for Ising spin glass and quenched disorder
models~\cite{berker09,kaufman82}, or disordered Potts models~\cite{igloi09}.
Multicritical point locations can moreover be checked carefully using Nishimori
symmetric lines~\cite{nishimori80,nobre01,robinson11} and duality exact
properties~\cite{nishimori81}. Frustration on hierarchical lattices has been
studied for example on a diamond shape ferromagnetic
structure with an additional transverse antiferromagnetic coupling between two
spins~\cite{oguchi09}. Important properties
of the phase diagram when the coupling is increased concern the low temperature
entropy which presents steps at specific values of this coupling reflecting
the frustration character of the lattice geometry. These transitions are
generated by slight variations of the transverse coupling inducing a change in
the nature of the ground states. Recursion relations can
be found exactly in this non-disordered but frustrated model, hence equations
for ground state structure, residual entropy and magnetization can be obtained
using scaling properties of the hierarchical geometry.

Experimental realizations of frustrated magnetic structures with spin dilution
can be found in spin-ice materials doped with rare earth elements, such as
Dy$_{2-x}$Lu$_x$Ti$_2$O$_7$, Dy$_{2-x}$Y$_x$Ti$_2$O$_7$, and
Ho$_{2-x}$Y$_x$T$_2$iO$_7$~\cite{chang10}, where $x$ parametrizes the
randomness~\cite{ke07}. These compounds are obtained from primary materials
Dy$_{2}$Ti$_2$O$_7$ or Ho$_{2}$Ti$_2$O$_7$, where magnetic ions Dy$^{3+}$ and
Ho$^{3+}$ occupy
the sites of a pyrochlore lattice made of tetrahedra connected to each other by
their vertices. The first magnetic rule for each
individual tetrahedron is given by 2 spins out of the tetrahedron and 2 spins
in, known as first ice rule~\cite{gingras11}. These magnetic ions are
then replaced by non-magnetic atoms Y or Lu, giving rise to an experimental
system of site dilution where a macroscopic fraction of sites is non
magnetic. One of the interesting physical properties of these systems is the
non-monotonic behavior of the zero-temperature entropy with respect with the
dilution level~\cite{ke07} from the measured specific heat, which can be
explained quite accurately within the Pauling approximation where tetrahedra are
treated as independent~\cite{ke07}.
Refined calculations with Monte-Carlo simulations on Husimi
cactus~\cite{gingras12} is a step beyond the model of Pauling which takes into
account
the correlations between tetrahedra on the pyrochlore lattice~\cite{jaubert}.
In the Bethe-Peierls approximation, a tree-like structure is
constructed from a
central site where the number of branches grows with the distance from the
central site. However in this geometrical approach, the branches never reconnect
together, a construction which forbids geometrical loops, but recursion
equations can be in general written in a systematic manner.
In this paper, we address the question of site disorder importance on thermal
properties in the same spirit as~\cite{oguchi09} where frustration plays an
important role.

We indeed consider explicitly an antiferromagnetic tetrahedron-like
hierarchical structure with Ising spins $\sigma_i=\pm 1$ located on the
vertices $i$. Disorder configurations are given by a set of additional
quenched random variables $\epsilon_i=0,1$, with probability
$x$ and $(1-x)$ respectively, located on the same sites (vertices) $i$, as shown
in \efig{fig_rec}. The replicative procedure to construct iteratively the
lattice is implemented starting from one single antiferromagnetic bond at level
$r=0$, until level $r>0$ is reached, as exemplified on \efig{fig_rec} for the
first two steps. Such structure is known to have a well defined thermodynamical
limit, when $r$ is large, for the free energy per
site~\cite{griffiths82,kaufman84}.
The zero temperature entropy of the tetrahedron-like structure at level $r=1$
can be evaluated exactly as we will see in the next sections and it appears to
be non-monotonic as function of the dilution.
We can then generalize the calculation for larger structures $r>1$ using
recursion
equations in the replica space, and obtain quantitative information about
frustration and disorder effects in the thermodynamical limit. This simple model
can be
considered as a good example on how to characterize spin-ice or spin-liquid
states and how these states may emerge for example from the
specific heat or susceptibility measurements.

\section{Notations and method}

\begin{figure}
\centering
\includegraphics[scale=0.8,clip]{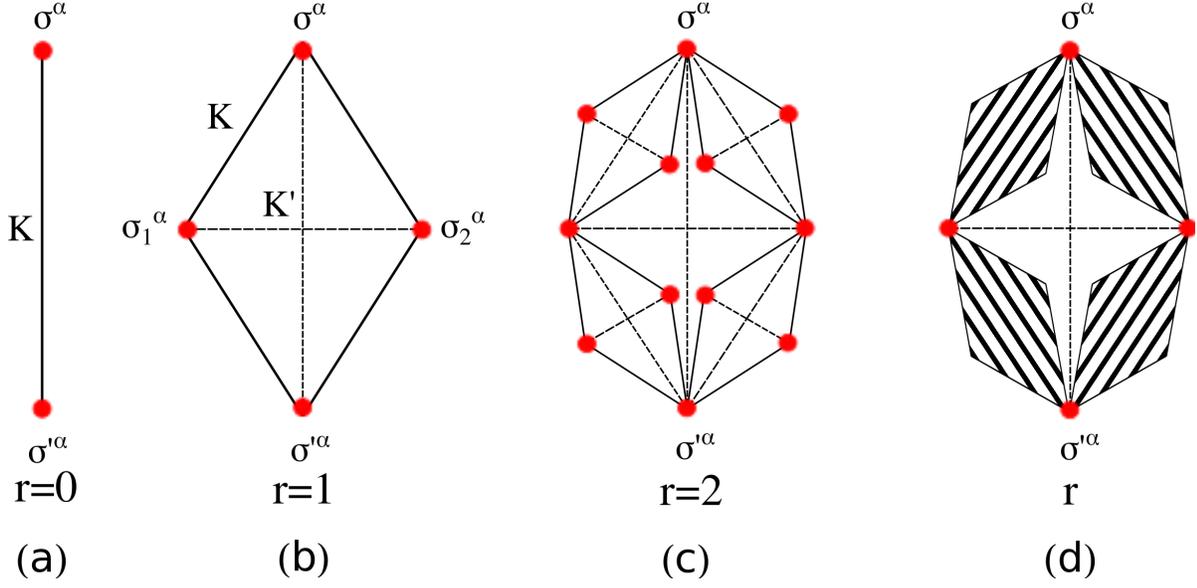}
\caption{\label{fig_rec}Construction of the hierarchical lattice, starting from
a single link $K>0$ between two spins (a). This link is replicated to form a
diamond shape, and couplings $\ac>0$ are added successively between
boundary sites, see (b) and (c), to form a frustrated lattice. Shaded areas in
(d) defines the previous lattice structures at level $r-1$, with bonds $\ac$
added furthermore between the boundary sites.}
\end{figure}

We consider the hierarchical lattice seen as a graph construction (set of
vertices connected here by links or edges), build recursively an arbitrary
number of times. Vertices can then be occupied by a spin with
probability $(1-x)$, with $0\le x\le 1$, or stays vacant with probability $x$
which is the dilution factor. Initially, at step $r=0$, the graph consists of a
single antiferromagnetic link, as displayed in \efig{fig_rec}(a), with two
vertices occupied eventually by spins.
This link has a Boltzmann strength $K=J/T$ with coupling unity $J=1$. The
structure at the next step is formed by replicating the previous structure four
times, and by connecting the lower and upper vertices together, as in
\efig{fig_rec}(b), (c) and in general (d). The diamond shape structures that are
obtained are connected two by two to form a larger graph. Additionally,
two transverse antiferromagnetic bonds of value $\ac>0$ [see \efig{fig_rec} (b)
(c) and (d)] are added at each step, in order to enhance frustration and
to mimic tetrahedral structures. For example, step $r=1$ is obtained by
connecting four single edges to form a square, on which two
antiferromagnetic edges of value $\ac$ are added transversely and
longitudinally.

The spins considered in this paper are Ising spins $\sigma_i=\pm 1$ on each
site (vertex) $i$, and we use an additional set of variables $\epsilon_i=0$
when the spin is vacant, with probability $x$, or $\epsilon_i=1$ when it is
present with probability $1-x$. It is straightforward to compute recursively the
total number of vertices $N_r$ at level $r$, equal to

\bb
N_0=2,\, N_1=4,\, N_{r+1}=4N_r-4, \, N_r=\frac{4^{r+1}+8}{6}.
\ee

Also the total number of bonds $B_r$ is equal to $B_r=4^r+2^{r+1}-2$, and the
maximum number of bonds connecting the two upper and lower sites is equal to
$L_r=2^r$. We may define an effective dimension $d$ such that $B_r=L_r^{d}$,
which gives $d=2$ in the large
size limit $r\gg 1$. In the following we will mainly concentrate on the
frustrated case $\ac=K$.
Partition function for this dilute hierarchical lattice made of
antiferromagnetic bonds can be constructed
starting at step $r=0$ using the following standard method. Also an
external magnetic field $h_e$ is taken into account in order to evaluate the
spin susceptibility and magnetization.

After $r$ recurrences, we define the partial partition function
$Z_r^{\{\eta\}}(\epsilon\sigma,\epsilon'\sigma')$ which
depends on some disorder configuration $\{\eta\}$ given
by a set of vacancy distribution $\{\epsilon_i=0,1\}_i$ with $i$
corresponding to the sites located inside
the structure delimited by the two extreme sites
$(\epsilon\sigma,\epsilon'\sigma')$. These latter sites have a given
disorder configuration $\{\epsilon,\epsilon'\}$ which does not belong
to $\{\eta\}$.
A recursion relation can be written between steps $r$ and $r+1$ of the lattice
construction

\bb\nn
\fl
Z_{r+1}^{\{\eta\}}(\epsilon\sigma,\epsilon'\sigma')&=&
\Tr{\sigma_1,\sigma_2}
Z_{r}^{\{\eta_1\}}(\epsilon\sigma,\epsilon_1\sigma_1)
Z_{r}^{\{\eta_2\}}(\epsilon_1\sigma_1,\epsilon'\sigma')
Z_{r}^{\{\eta_3\}}(\epsilon\sigma,\epsilon_2\sigma_2)
Z_{r}^{\{\eta_4\}}(\epsilon_2\sigma_2,\epsilon'\sigma')
\\ \label{Zdef}
&\times
&\exp(-\ac\epsilon_1\sigma_1\epsilon_2\sigma_2+\he(\epsilon_1\sigma_1+\epsilon_2
\sigma_2))
\times \exp(-\ac\epsilon\sigma\epsilon'\sigma')
2^{\epsilon_1-1}2^{\epsilon_2-1}
\ee

where $\{\eta\}$ represents the disorder configuration
$\{\eta\}=\{\eta_1,\eta_2,\eta_3,\eta_4,\epsilon_1,\epsilon_2\}$
and the $\Tr{.}$ sign the sum over all the possible spin values.
The additional factors $2^{\epsilon_1-1}2^{\epsilon_2-1}$ compensate the fact
that we sum over
eventual ghost spins located on vacant sites.
The initial condition \efig{fig_rec}(a) is given by the two-spin partition
function

\bb
Z_0^{\{\eta\}}(\epsilon\sigma,
\epsilon'\sigma')=\exp(-K\epsilon\epsilon'\sigma\sigma')
\ee

which has actually no internal disorder dependence, $\{\eta\}=\{\emptyset\}$.
The magnetic field is implemented in \eref{Zdef} only for spins not belonging
to the top and bottom vertices.
Using replica method, we consider $n$ copies of \eref{Zdef} for a
given configuration $\{\eta\}$ and perform the average

\bb
Z_{r+1}(\epsilon\sigma^{\alpha},\epsilon'\sigma'^{\alpha}):=
\Sp{\prod_{\alpha=1}^n
Z_{r+1}^{\{\eta\}}(\epsilon\sigma^{\alpha},\epsilon'\sigma'^{\alpha})}{\eta}.
\ee

Here $[\,.\,]_{\eta}$ is meant for averaging over disorder $\{\eta\}$. In this
case, the recursion relation \eref{Zdef} becomes

\bb\nn\fl
Z_{r+1}(\epsilon\sigma^{\alpha},\epsilon'\sigma^{\alpha'})&=&
\exp(-\ac \epsilon\epsilon'\sum_{\alpha}\sigma^{\alpha}\sigma'^{\alpha})
\int P(\epsilon_1)P(\epsilon_2)d\epsilon_1d\epsilon_2
\Tr{\sga,\sgb}
\\ \nn
& &
\times
Z_{r}(\epsilon\sigma^{\alpha},\epsilon_1\sga)Z_{r}(\epsilon_1\sga,
\epsilon'\sigma'^{\alpha})
Z_{r}(\epsilon\sigma^{\alpha},\epsilon_2\sgb)Z_{r}(\epsilon_2\sgb,
\epsilon'\sigma'^{\alpha})
\\ \label{eq_rec}
& &\times \exp\Big [-\ac
\epsilon_1\epsilon_2\sum_{\alpha}\sga\sgb+\he\sum_{\alpha}(\sga+\sgb)\Big ]
2^{n(\epsilon_1-1)}2^{n(\epsilon_2-1)}
\ee

where $P$ is the bimodal distribution for the dilute sites
$P(\epsilon):=x\delta(\epsilon)+(1-x)\delta(\epsilon-1)$. In the previous
expression, we perform only the sum over the two spins
$\sga$ and $\sgb$ connecting pairs of diamond shape structures. Starting
with initial condition

\bb
Z_0(\epsilon\sigma^{\alpha},\epsilon'\sigma^{\alpha'}
)=\exp(-K\epsilon\epsilon'\sum_{\alpha}\sigma^{\alpha}
\sigma'^{\alpha})
\ee

all quantities $Z_r(\epsilon\sigma^{\alpha},\epsilon'\sigma^{\alpha'})$ can
in principle be computed step by step. Free energy $\free{r}$ is evaluated
directly after averaging over the remaining disorder and
summation over the two spin variables. We define the complete partition
function as

\bb\label{zrn}\fl
z_r(n)=\int P(\epsilon)P(\epsilon')d\epsilon
d\epsilon'2^{n(\epsilon-1)}2^{n(\epsilon'-1)}
\Tr{\sigma^{\alpha},\sigma'^{\alpha}}Z_r(\epsilon\sigma^{\alpha},
\epsilon'\sigma^{\alpha'})
\exp\Big [\he\sum_{\alpha}(\sigma^{\alpha}+\sigma'^{\alpha})\Big ]
\ee

and take the limit $n\rightarrow 0$

\bb
-K \free{r}:=\lim_{n\rightarrow 0}\frac{1}{n}\left ( z_r(n)-1 \right )=z_r'(0)
\ee

after noticing that $z_r(0)=1$. The disorder averaged entropy can be expressed
in terms of $z_r'(0)$ via the usual thermodynamical relations

\bb\label{Sent}
\ent{r}(T):=-\frac{\partial \free{r}}{\partial
T}=z_r'(0)-K\frac{\partial}{\partial K}z_r'(0)
\ee

as well as the averaged specific heat $\cv{r}=-T\partial^2 \free{r}/\partial
T^2=K^2\partial^2 z_r'(0)/\partial K^2$.
The relation between entropy and specific heat is given by the integral

\bb
\ent{r}(T)=N_r(1-x)\ln(2)-\int_T^{\infty}\frac{\cv{r}}{T'}dT'
\ee

which is a direct experimental way to measure the zero-temperature entropy by
extrapolation, knowing that all spins are in the paramagnetic phase at high
temperature, each contributing with a factor $\ln(2)$.
The linear susceptibility corresponds to the excitation of the magnetic order
parameter $M=\sum_i \sigma_i$ under a field $h_e$ and is defined as
$\suscept{r}=\partial [<M>]_{\eta}/\partial h_e$, after averaging over disorder.
We can also relate $\suscept{r}$ to the free energy, using
$T\suscept{r}=[<M^2>]_{\eta}-[<M>^2]_{\eta}$ and $[<M>]_{\eta}=-\partial
\free{r}/\partial h_e$, in particular

\bb\label{susc}
\suscept{r}=\frac{\partial [<M>]_{\eta}}{\partial h_e}=-\frac{\partial^2
\free{r}}{\partial h_\e^2}.
%\,\,\chi_3=\frac{1}{6}\frac{\partial^4 F_r}{\partial h_\e^4}.
\ee

\section{Residual entropy for $r=1$ structure}

As a simple application, we consider the case $r=1$ consisting in only 4
spins \efig{fig_rec}(b), where all thermodynamical quantities can be computed
exactly. The zero temperature entropy is identical to the entropy of a
three-dimensional tetrahedron made of antiferromagnetic bonds. After iterating
\eref{eq_rec}, we obtain

\bb\nn\fl
& &Z_1(\epsilon\sigma^{\alpha},\epsilon'\sigma'^{\alpha})=
\exp(-\ac \epsilon\epsilon'\sum_{\alpha}\sigma^{\alpha}
\sigma'^{\alpha})\times
\\ \nn
& &\left [
(1-x)^22^n\prod_{\alpha=1}^n\left (\e^{\ac }+\e^{-\ac }\cosh \Big [
2K(\epsilon\sigma^{\alpha}+\epsilon'\sigma'^{\alpha})-2\he\Big ] \right )
\right .
\\
& &\left .+2x(1-x)2^n\prod_{\alpha=1}^n\cosh\Big [ K(\epsilon\sigma^{\alpha}
+\epsilon'\sigma'^{\alpha})-\he\Big ]+x^2 \right ].
\ee

Then the function $z_1(n)$ is equal, after some computation, to

\bb\nn
z_1(n)&=&(1-x)^44^n\left [
1+\Big \{\cosh(4K)+\e^{-4K}\sinh^2(2\he)\Big \}\e^{-2\ac
}+\e^{2\ac}+\cosh(2\he)\right ]^n
\\ \nn
&+&4x(1-x)^3 4^n \left [
\cosh(\he)\e^{\ac }+\Big \{\cosh(2 K)\cosh(\he)\cosh(2\he) \right .
\\ \nn
&-&\left .\sinh(2 K)\sinh(\he)\sinh(2\he)\Big \}\e^{-\ac } \right ]^n
\\ \nn
&+&2x^2(1-x)^22^n\left [
\Big (\cosh(2\he)\e^{-\ac}+\e^{\ac}\Big )^n
+2\Big (\cosh(2\he)\e^{-K}+\e^{K}\Big )^n \right ]
\\ \label{z1}
&+&4x^3(1-x)2^n\cosh^n(\he)
+x^4.
\ee

When $K$ is large and in absence of external field, this expression can be put
into the following form, by keeping
the dominant contributions in the exponential terms

\bb
z_1(n)=\sum_{k\ge 0}\rho_k \e^{ n(s_k-Ke_k)},\;\sum_{k\ge 0}\rho_k=1.
\ee

This expansion is useful in order to identify the zero temperature entropy which
can be written as a sum of contributions $\ent{1}(T=0)=\sum_k\rho_ks_k$.
Each partial entropy $s_k$ and energy $e_k$ physically corresponds to a
configuration of disorder with weight $\rho_k$. In this limit, the asymptotic
form for $z_1(n)$  when $0\le \ac < K$ is indeed equal to

\bb\nn
z_1(n)&\simeq &(1-x)^42^n\e^{n(4K-2\ac)}+4x(1-x)^32^n\e^{n(2K-\ac)}
\\
&+&4x^2(1-x)^22^n\e^{nK}+2x^2(1-x)^22^n\e^{n\ac }+4x^3(1-x)2^n+x^4.
\ee

After taking the limit $n=0$, the entropy and energy per spin are respectively
equal to
$\ent{1}(T=0)/[4(1-x)]=(1-x^4)/[4(1-x)]\ln(2)$ and
$\en{1}/[4(1-x)]=-(1-x)^3(1-\acd/2)-x(1-x)^2(2-\acd)-x^2(1-x)(1+\acd/2)$,
where $\acd:=\ac/K$ is the coupling ratio.
The frustrated case $\ac=K$ is particular since we obtain instead

\bb\nn
z_1(n)&\simeq &(1-x)^46^n\e^{n2K}+4x(1-x)^36^n\e^{nK}+
6x^2(1-x)^22^n\e^{nK}
\\ \label{eq_z1}
&+&4x^3(1-x)2^n+x^4
\ee

and, in this specific case, the entropy and energy per spin have a different
expression

\bb\nn
\frac{\ent{1}(T=0)}{4(1-x)}&=&(1-x)^3\frac{\ln(6)}{4}+x(1-x)^2\ln(6)
+3x^2(1-x)\frac{\ln(2)}{2}
+x^3\ln(2),
\\ \label{S1}
\frac{\en{1}(T=0)}{4(1-x)}&=&-\ff(1-x)^3-x(1-x)^2-\frac{3}{2}x^2(1-x).
\ee

As expected, the entropy for a single tetrahedron in the frustrated
case is larger. Plot of the entropy is given in \efig{fig_SvsxRes} as function
of $x$ and appears to be
non-monotonic with a local maximum around $x=0.295$. This maximum, intrisic to
a 4 spin system or single uncoupled tetrahedra, is probably due to the
non-monotonic variation of the fraction of the spin configurations with the
lowest energy as function of dilution in a system of 4 spins where there are at
most 16 possible states, in a similar way as it is describes in~\cite{ke07} for
coupled tetrahedra. This maximum disappears for $r>1$ as shown on the same
figure because the varying connectivity and coupling arrangement on the
hierarchical structure tends to modify the fractions of these acceptable states
in the different tetrahedral units. The exact entropy value in
absence of dilution is given by $\ln(6)/4\simeq 0.4479$ which is larger than for
a system of coupled tetrahedra in the large size limit (see sections below).

\begin{figure}
\centering
\includegraphics[scale=0.5,clip]{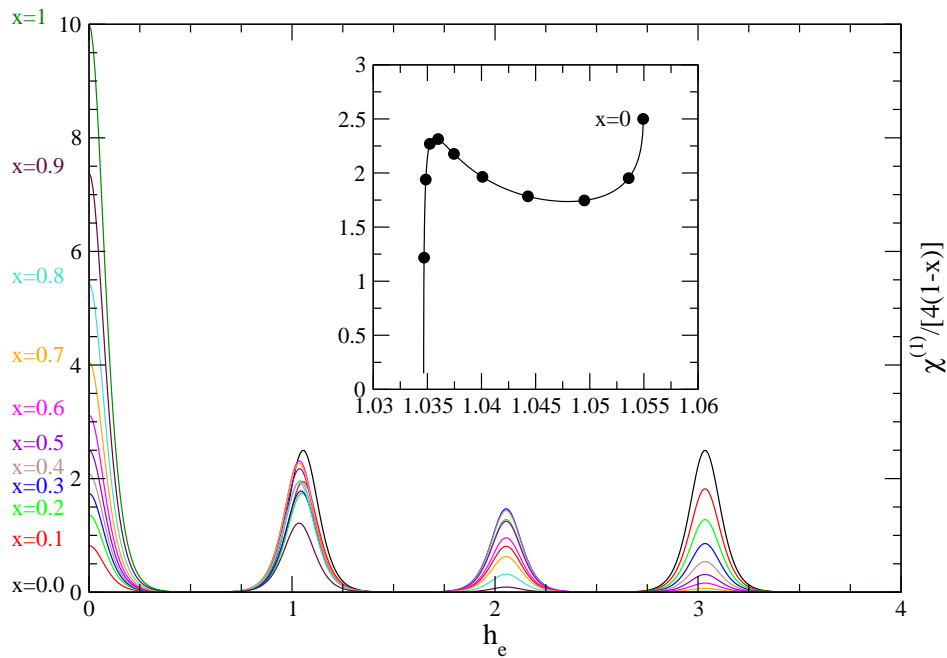}
\caption{\label{fig_chi4} (Color online) Susceptibility per spin
$\chi^{(1)}/[4(1-x)]$ at
$T=0.1$ as function of the external field $h_e$,
for different values of dilution factor. In absence of disorder (black
lines), the susceptibility is composed of two peaks located at $h_e=1,3$. Inset:
variation of the second peak position with dilution.}
\end{figure}%

Susceptibility $\chi^{(1)}$ is derived exactly using definition \eref{susc} and
is plotted in \efig{fig_chi4} as function of the
external field for different values of $x$. As expected, when $x>0$, a peak
appears at $h_e=0$ when the probability to
find a single spin in the tetrahedron is non zero. A simple calculation for the
non-disordered case leads to three possible ground states
per spin $E_G$ depending on the value of the field: two spins up and two spins
down with $E_G=-2$, three spins up and one spin down with $E_G=-2h_e$, and all
spins up with $E_G=6-4h_e$. Two transitions occur respectively at $h_e=1$ and
$h_e=3$.

\begin{figure}
\centering
\includegraphics[scale=0.5,clip]{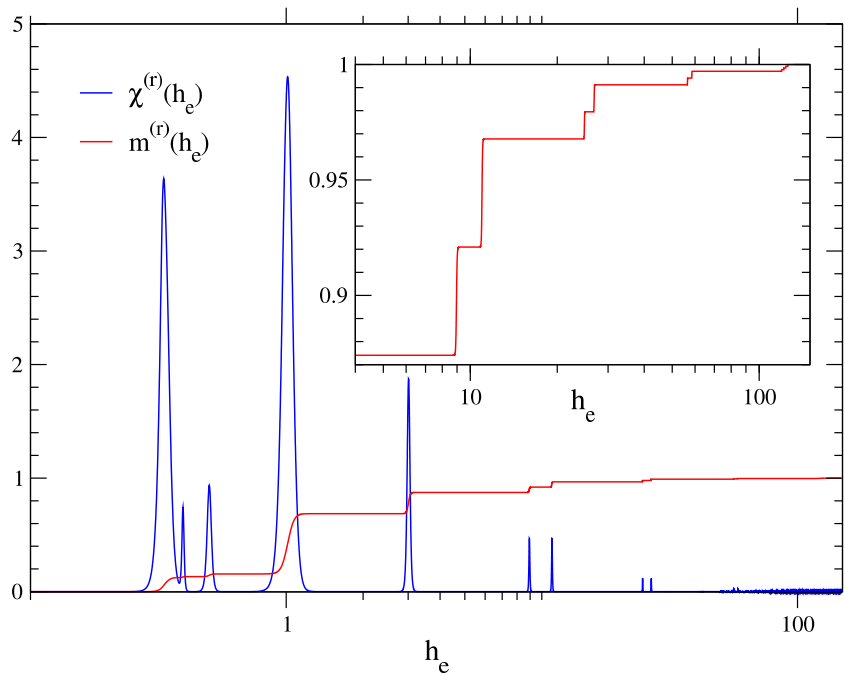}
\caption{\label{fig_chipure} (Color online) Susceptibility (blue line)
$\chi^{(r)}(h_e)$ and magnetization $m^{(r)}(h_e)$ (red line) per spin
at $T=0.05$ as function of the external field $h_e$ (logarithmic scale) in
absence of disorder for level $r=6$ (2732 sites).
Inset: magnetization per spin in the large field limit until saturation.
}
\end{figure}%

For larger sizes however, it is interesting to analyze the susceptibility using
recursive equations \eref{eq_rec} for the partition
function $Z_r(\sigma,\sigma')$ in the absence of disorder $x=0$. The recursive
equations are detailed in \ref{app1}. The method is
to assume an effective form for the partition function, with Ising
effective coupling, magnetic field, and weight coefficient at all recursion
levels. The exact recurrence equations for these three quantities as function
of the external magnetic field $h_e$ allow us directly to compute the
susceptibility. Plots of $\chi^{(r=6)}(h_e)$ and magnetization
per spin $m^{(r=6)}(h_e)$ are displayed in \efig{fig_chipure} showing multiple
field transitions and magnetization plateaus up to $h_e=127$ for $N_6=2732$
spins. In the next section, the analysis is extended to
the dilute case. We use the set of recursive equations \eref{eq_rec} to
compute the partition function at a given step $r$.
The main technique is to reorganize the partition function in the same manner as
\eref{eq_z1} where we easily identify weights $\rho_k=x^k(1-x)^{4-k}\bin{4}{k}$.
One difficulty with dilution is that an iterative procedure at low temperature
on effective Ising constant couplings to obtain the limiting coupling
distribution can in principle be performed, but not for the free energy,
unlike the case of spin glass models. For example, for the disordered Potts
model~\cite{igloi09} where the disorder is given by a discrete distribution of
couplings, zero temperature free energy satisfies some recursive equations.
In general, when the disorder is located only in the couplings, the four
structures represented as shaded areas in \efig{fig_rec}(c) have independent
internal disorder, even if they are connected together by the four vertices.
Therefore disorder can be factorized and direct iteration
of Ising couplings by renormalization can be performed as well as the partition
function weights such as $I_r$ in \eref{rec_x0} which are important for the
determination of free energy and entropy. In the case of dilution however, the
disorder configuration in the
boundary vertex sites given by set
$\{\epsilon,\epsilon',\epsilon_1,\epsilon_2\}$ is actually shared between the
four diamond structures. For example variable $\epsilon_1$ is common to the two
shaded diamonds on the left hand side of \efig{fig_rec}(c). This makes the
evaluation of weight coefficients $I_r$ \eref{rec_x0} problematic in the
disordered case since after few iterations correlations will develop rapidly.
This is main reason why it is more convenient to consider
a replica version of \eref{rec_x0} and \eref{Zdef} with partial integration over
the disorder located strictly inside the diamond structures, as
explicitly defined previously by \eref{eq_rec} . As we will see,
an approximation
scheme can be derived from the replica method and
thermodynamical functions can be obtained.

\section{Recursion relations in the general case}

At step $r$, we assume from the previous analysis an expansion of the
partition function in terms of configurational weights
$x^k(1-x)^{n_r-k}\bin{n_r}{k}$ where $n_r$ is here the number of sites localized
between the two extreme top and bottom sites and which satisfies the equation
$n_{r+1}=4n_r+2$, with initial condition $n_0=0$ and $N_r=n_r+2$.
The main idea, as discussed before, is to keep a generic and minimal expression
for the partial partition function, which depends on unknown functions
satisfying recurrence equations such that

\bb\nn
Z_r(\epsilon\sigma^{\alpha},\epsilon'\sigma'^{\alpha})&=&
\exp(-\ac \epsilon\epsilon'\sum_{\alpha}\sigma^{\alpha}
\sigma'^{\alpha})\sum_{k=0}^{n_r}x^k(1-x)^{n_r-k}
\bin{n_r}{k}\exp\left (
nI_k^{(r)}(\epsilon,\epsilon')
\right .
\\ \label{expand}
&+&\left .
K_k^{(r)}\epsilon\epsilon'\sum_{\alpha}\sigma^{\alpha}\sigma'^{\alpha}+H_k^{(r)}
(\epsilon,\epsilon')
\sum_{\alpha}[\epsilon\sigma^{\alpha}+\epsilon'\sigma'^{\alpha}]
\right )
\ee

where $H_k^{(r)}(\epsilon,\epsilon')$ and $I_k^{(r)}(\epsilon,\epsilon')$ are
symmetric functions of
$\epsilon$ and $\epsilon'$. Coupling $K_k^{(r)}$ is independent of the boundary
disorder. All these functions
depend implicitly on $K$ and $h_e$. As initial condition we impose
$K_0^{(0)}=-K+\ac$
and $H_0^{(0)}(\epsilon,\epsilon')=I_0^{(0)}(\epsilon,\epsilon')=0$. We also
take $\ac=K$ in the following for the frustrated
version. At the next level,  $Z_{r+1}$ is evaluated by considering the product
of four partition functions $Z_r$ as written in \eref{eq_rec}.

Using the ansatz \eref{expand}, this product is expanded as

\bb\nn\fl
Z_{r+1}=\exp(-\ac
\epsilon\epsilon'\sum_{\alpha}\sigma^{\alpha}\sigma'^{\alpha})
\sum_{k_1,k_2,k_3,k_4=0}^{n_r}
\bin{n_r}{k_1}\bin{n_r}{k_2}\bin{n_r}{k_3}\bin{n_r}{k_4}
x^{k_1+k_2+k_3+k_4}(1-x)^{4n_r-k_1-k_2-k_3-k_4}
\\ \nn\fl
\times\int P(\epsilon_1)P(\epsilon_2)d\epsilon_1d\epsilon_2\Tr{\sga,\sgb}
\exp\left (
nI_{k_1}^{(r)}(\epsilon,\epsilon_1)+nI_{k_2}^{(r)}(\epsilon_1,\epsilon')
+nI_{k_3}^{(r)}(\epsilon,\epsilon_2)+nI_{k_4}^{(r)}(\epsilon_2,\epsilon')
\right )
\\ \nn\fl
\times\exp \left (
-\ac \epsilon_1\epsilon_2\sum_{\alpha}\sga\sgb
-\ac
\sum_{\alpha}(\epsilon_1\sga+\epsilon_2\sgb)(\epsilon\sigma^{\alpha}
+\epsilon'\sigma'^{\alpha})
\right ) 2^{n(\epsilon_1-1)}2^{n(\epsilon_2-1)}
\\ \nn\fl
\times\exp\left (
K_{k_1}^{(r)}\epsilon\epsilon_1
\sum_{\alpha}\sigma^{\alpha}\sigma_1^{\alpha}
+K_{k_2}^{(r)}\epsilon_1\epsilon'
\sum_{\alpha}\sigma_1^{\alpha}\sigma'^{\alpha}
+K_{k_3}^{(r)}\epsilon\epsilon_2
\sum_{\alpha}\sigma^{\alpha}\sigma_2^{\alpha}
+K_{k_4}^{(r)}\epsilon_2\epsilon'
\sum_{\alpha}\sigma_2^{\alpha}\sigma'^{\alpha}
\right )
\\ \nn\fl
\times\exp\left (
H_{k_1}^{(r)}\sum_{\alpha}[\epsilon\sigma^{\alpha}+\epsilon_1\sigma_1^{\alpha}]
+H_{k_2}^{(r)}\sum_{\alpha}[\epsilon'\sigma'^{\alpha}+\epsilon_1\sigma_1^{\alpha
}]
+H_{k_3}^{(r)}\sum_{\alpha}[\epsilon\sigma^{\alpha}+\epsilon_2\sigma_2^{\alpha}]
+H_{k_4}^{(r)}\sum_{\alpha}[\epsilon'\sigma'^{\alpha}+\epsilon_2\sigma_2^{\alpha
}]
\right )
\\ \label{rec_prod}
\times
\exp\Big [\he\sum_{\alpha}(\epsilon_1\sga+\epsilon_2\sgb)\Big ].
\ee

The last term takes into account the missing field on former boundary spins
$\sga$ and $\sgb$ which are now summed up.
It is useful to introduce the operator
$1=\sum_{k=0}^{4n_r}\delta_{k,k_1+k_2+k_3+k4}$ or the integral form

\bb\label{unit}
1=\sum_{k=0}^{4n_r}\delta_{k,k_1+k_2+k_3+k4}=\sum_{k=0}^{4n_r}
\int_{0}^{2\pi}\frac{d\theta}{2\pi} {\rm e}^{ i\theta(-k+k_1+k_2+k_3+k_4)}
\ee

which is then inserted in the previous expression in order to reorganize the sum
over the $k_i$ into a single sum over
weights $x^k(1-x)^{4n_r-k}\bin{4n_r}{k}$. The integral over $\theta$ can be
performed using a first order expansion
in $n$, sufficient to obtain $z_r'(0)$. For example, given integer $p>1$ and
$0\le k\le pn_r$, let us consider $n_r+1$ field variables $\varphi_l$
and define the quantity $W_{n,k}[\varphi_l]$ made of the product of $p$ sums
$\sum_{k_i}\bin{n_r}{k_i}\exp(n\varphi_{k_i})$, $i=1,\cdots,p$ by analogy with
the product of sums that appears in \eref{rec_prod}. In addition, we impose the
constraint $\sum_{i=1}^{p}k_i=k$ by using the Kronecker integral \eref{unit},
and perform an expansion at first order in $n$

\bb\label{identity}\fl
W_{n,k}[\varphi_{l}]:=\int_{0}^{2\pi}\frac{d\theta}{2\pi} {\rm e}^{-i\theta k}
\prod_{i=1}^p
\left (
\sum_{k_i=0}^{n_r}
\bin{n_r}{k_i}{\rm e}^{ i\theta k_i+n\varphi_{k_i}}
\right )
\\ \nn
\simeq
\int_{0}^{2\pi}\frac{d\theta}{2\pi} {\rm e}^{-i\theta k}
\prod_{i=1}^p
\left [
(1+{\rm e}^{ i\theta})^{n_r}
+n\sum_{k_i}\bin{n_r}{k_i}{\rm e}^{ i\theta k_i}\varphi_{k_i}
\right ]
\\ \nn
\simeq
\int_{0}^{2\pi}\frac{d\theta}{2\pi} {\rm e}^{-i\theta k}
(1+{\rm e}^{ i\theta})^{pn_r}
\left [
1+np(1+{\rm e}^{ i\theta})^{-n_r}\sum_{k_1}\bin{n_r}{k_1}{\rm e}^{ i\theta
k_1}\varphi_{k_1}
\right ]
\\ \nn
=\bin{pn_r}{k}+np\sum_{k_1}\bin{n_r}{k_1}\bin{(p-1)n_r}{k-k_1}
\varphi_{k_1}\simeq
\bin{pn_r}{k}\exp\left
(np\sum_{k_1=\max(0,k-(p-1)n_r)}^{\min(k,n_r)}\frac{\bin{n_r}{k_1}\bin{(p-1)n_r}
{k-k_1}}{\bin{pn_r}{k}}
\varphi_{k_1}\right ).
\ee

The exponentiation in the last line allows us, at first order in $n$, to
reorganize the product of the $p=4$
sums in \eref{eq_rec} as a single sum over configurations of $k$ vacant sites,
with combinatorial
factor $\bin{pn_r}{k}$. In particular, taking a constant value
$\varphi_{l}:=\varphi$, we easily find
$W_{n,k}[\varphi]=\bin{pn_r}{k}\e^{np\varphi}$, and therefore the identity

\bb\label{norm}
\sum_{k_1=\max(0,k-(p-1)n_r)}^{\min(k,n_r)}\frac{\bin{n_r}{k_1}\bin{(p-1)n_r}{
k-k_1}}{\bin{pn_r}{k}}=1
\ee

from which we deduce that factors

\bb
{\cal
D}^{k,k_1}_{n_r,p}:=\frac{\bin{n_r}{k_1}\bin{(p-1)n_r}{k-k_1}}{\bin{pn_r}{k}}
\ee

can be considered as natural weights since the sum over integers $k_1$ is
normalized. We may apply
this technique to spin operators $\sum_{\alpha}\sigma^{\alpha}\sigma'^{\alpha}$
appearing in \eref{rec_prod}
as well, which are sum of $n$ terms. Considering for example linear spin
operator
$\sum_{\alpha}\sigma^{\alpha}$, and instead of \eref{identity} the function

\bb\label{identity_op}
W_{n,k}[\varphi_{l}]&:=&\Tr{\sigma^{\alpha}}\int_{0}^{2\pi}\frac{d\theta}{2\pi}
{\rm e}^{-i\theta k}
\prod_{i=1}^p
\left (
\sum_{k_i=0}^{n_r}
\bin{n_r}{k_i}{\rm e}^{ i\theta k_i+\varphi_{k_i}K\sum_{\alpha}\sigma^{\alpha}}
\right )
\\ \nn
&=&\int_{0}^{2\pi}\frac{d\theta}{2\pi} {\rm e}^{-i\theta k}
\sum_{\{k_i\}}\left (\prod_{i=1}^p
\bin{n_r}{k_i}\right ){\rm e}^{ i\theta\sum_i k_i}
2^n\cosh^n(K\sum_i\varphi_{k_i}).
\ee

We then take the limit $n\rightarrow 0$ and define $w_k[\varphi_{l}]:=\partial
W_{n,k}[\varphi_{l}]/\partial n|_{n=0}$, so that

\bb\label{wfunc}
w_k[\varphi_{l}]=
%\int_{0}^{2\pi}\frac{d\theta}{2\pi} {\rm e}^{-i\theta k}
\sum_{\sum_i k_i=k}\left (\prod_{i=1}^p
\bin{n_r}{k_i}\right )
%{\rm e}^{ i\theta\sum_i k_i}
\ln\left [2\cosh(K\sum_i\varphi_{k_i})\right ].
\ee

We can compare this expression with the approximation

\bb\label{wfuncapp}
w_k[\varphi_{l}]\simeq \tw_k[\varphi_{l}]:= \bin{pn_r}{k}\ln\left
[2\cosh\Big (pK\sum_{k_1}{\cal
D}^{k,k_1}_{n_r,p}\varphi_{k_1}\Big )\right ]
\ee

coming from the same analysis made in \eref{identity}, we can discuss two
different cases.
First, let choose $\varphi_l:=\varphi$, both functions $w_k$ and
$\tw_k$ are identical, using the
normalization \eref{norm}. Then, we may try non-constant fields such as
$\varphi_{l}:=l\varphi$, which
gives after summation the exact result

\begin{figure}
\centering
\includegraphics[scale=0.5,clip]{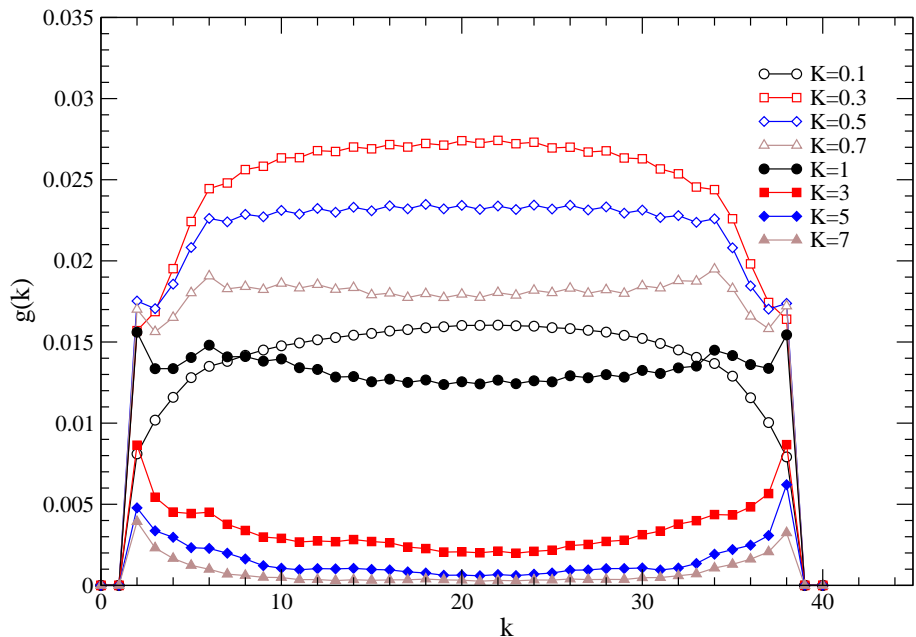}
\caption{\label{fig_test}(Color Online)
Error function $g(k)$ (see text for definition) that measures the accuracy
of the approximation \eref{identity} applied to function
\eref{identity_op} and performed on random variables $\varphi_l$
with a Poisson distribution. Here $n_r=10$ and $0\le k\le 4n_r$, and 15000
realizations were performed before averaging.
}
\end{figure}

\begin{figure}
\centering
\includegraphics[scale=0.5,clip]{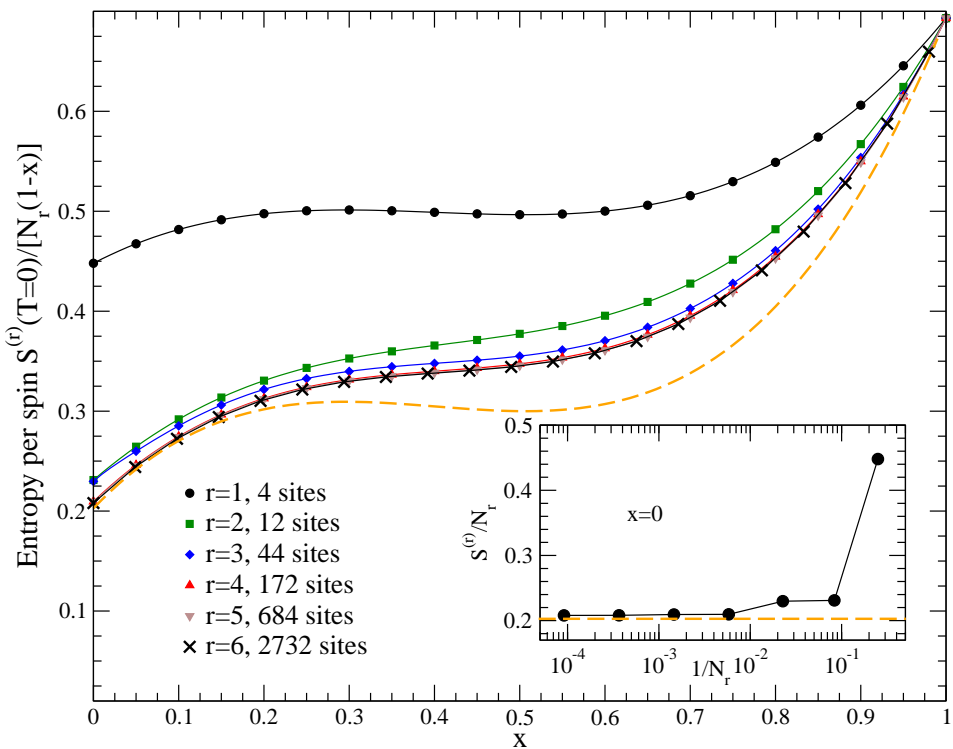}
\caption{\label{fig_SvsxRes}(Color Online) Residual entropy per spin
$\ent{r}/[N_r(1-x)]$ at zero temperature computed from the expression of the
free energy \eref{freeFr}. The entropy limit per spin for a very
dilute system (independent spins) is close to $\ln(2)=0.6931$ as expected. The
orange dashed line represents the Pauling residual
entropy $S_P(x)$ given in \eref{SP}, with a value for the undilute case equal
to $S_P(x=0)=\ff\ln(3/2)\simeq 0.2027$ (see text).
Inset: zero temperature entropy for the undilute system as function of the
inverse system size.
First exact values are $\ent{1}/4=\ln(6)/4=0.4479$,
$\ent{2}/12=\ln(2)/3=0.2310$, $\ent{3}/44\simeq 0.2297$. The term at recursion
level 6 is approximately
equal to $\ent{6}/2732\simeq 0.20804$. The dashed straight line is the Pauling
entropy $S_P(0)$.}
\end{figure}

\bb
w_k[\varphi_{l}=l\varphi]=\bin{pn_r}{k}\ln\Big [2\cosh(kK\varphi)\Big ],
\ee

which is also identical to $\tw_k$ in \eref{wfuncapp} using equality
$\sum_{k_1}{\cal D}^{k,k_1}_{n_r,p}k_1=k/p$. In the more general case, when
the arguments $\phi_{l\le n_r}$ are random variables, we can try to evaluate the
accuracy of \eref{wfuncapp}. Let us consider for example a Poisson distribution
for the $\varphi_l>0$, with mean and variance unity,
$\rm{prob}(\varphi_l)=\exp(-\varphi_l)$, and $K$ fixed.
A measure of the accuracy can be given by the relative error function
$g(k):=<(\tw_k[\varphi_l]/w_k[\varphi_l]-1)^2>^{1/2}$, where the brackets
are the average over random realizations. In \efig{fig_test} is represented
$g(k)$ as function of $k$ for $n=10$ and for different temperatures $1/K$.

%--------------------------------------------
%---------------------------------------------

Applying \eref{identity} for spin operators will be useful to simplify
expression \eref{rec_prod} and obtain recursive equations
in presence of dilution.
Computing recursive equations for $H_k^{(r)}(\epsilon,\epsilon')$,
$I_k^{(r)}(\epsilon,\epsilon')$, and $K_k^{(r)}$
follows two steps: the integral over variable $\theta$ coming from the
constraint
$k_1+k_2+k_3+k_4=k$, with $k_i=0,\cdots,n_r$, introduced in \eref{rec_prod}, is
performed
using transformation \eref{identity} over the different spin operators with
$p=4$.
This will allow for the partial summation over spins $\sga$ and $\sgb$, and
averaging over random variables $\epsilon_1$ and
$\epsilon_2$, see \efig{fig_rec}(c). We obtain an expression for $Z_{r+1}$ as a
summation over configurations
$x^{k+l}(1-x)^{4n_r+2-k-l}\bin{4n_r}{k}\bin{2}{l}$, with
$4n_r$ sites coming from inside the shaded diamond structures in
\efig{fig_rec}(c), plus the two sites coming from the integration over
$\epsilon_1$ and $\epsilon_2$. This sum can be furthermore reorganized using
again identity \eref{identity} in order to finally obtain
\eref{expand} at level $r+1$ as a sum over weights
$x^k(1-x)^{4n_r+2-k}\bin{4n_r+2}{k}$ (with $4n_r+2=n_{r+1}$) and new
coupling values.
All details of this development are presented in \ref{app2}, where
recurrence equations are written explicitly in \eref{newcouplings}.

Using ansatz \eref{expand}, and integrating over the boundary site degrees of
freedom, the complete partition function \eref{zrn} is then equal to

\bb\nn\fl
z_r(n)=
\sum_{k=0}^{n_r}\bin{n_r}{k}x^{k}(1-x)^{n_r-k}
\Big [ (1-x)^2
\exp(nI^{(r)}_k(1,1))2^n
\Big \{\e^{K_k^{(r)}-\ac}\cosh\Big [2H_k^{(r)}(1,1)+2\he\Big
]+\e^{-K_k^{(r)}+\ac}\Big \}^n
\\ \nn
+2x(1-x)\exp(nI^{(r)}_k(0,1))2^n\cosh^n\Big [H_k^{(r)}(0,1)+\he\Big ]+x^2
\exp(nI^{(r)}_k(0,0)) \Big ]
\ee

and the free energy is derived directly from the previous equation

\bb\nn\fl
-K\free{r}=z_r'(0)=
\sum_{k=0}^{n_r}\bin{n_r}{k}x^{k}(1-x)^{n_r-k}
\Big [ (1-x)^2\Big \{\ln(2)+I^{(r)}_k(1,1)
\\ \label{freeFr}
+\ln\Big (\e^{K_k^{(r)}-\ac}\cosh\Big [2H_k^{(r)}(1,1)+2\he\Big
]+\e^{-K_k^{(r)}+\ac}\Big )
\Big \}
\\ \nn
+2x(1-x)\Big \{\ln(2)+I^{(r)}_k(0,1)+
\ln\cosh\Big [H_k^{(r)}(0,1)+\he\Big ]
\Big \}+x^2I^{(r)}_k(0,0) \Big ].
\ee

After a few steps, the number of configurations is growing rapidly, as the
number of weights  $\bin{n_r}{k}x^{k}(1-x)^{n_r-k}$
becomes exponentially large as well as the number of iterative functions to
evaluate. We can obtain however a very good approximation is we notice that
these weights are distributed closely around a Gaussian when $n_r$ is
sufficiently large

\bb
\bin{n_r}{k}x^{k}(1-x)^{n_r-k}\simeq
%\frac{\e^{-(k-xn_r-x+\ff)^2/[2x(1-x)n_r]}}{\sqrt{2\pi x(1-x)n_r}}.
\frac{\exp\Big (-\frac{(k-xn_r-x+\ff)^2}{2x(1-x)n_r}\Big )}{\sqrt{2\pi
x(1-x)n_r}}
\ee

For $r=4$ for example, the number of internal sites is equal to $n_r=170$, and
the previous approximation is very accurate.
Numerically, we solved the iterative functions up to level $r=4$ included, using
\eref{newcouplings}, and then apply
for higher levels $r>4$ the Gaussian approximation for $k$ distributed with 4
standard deviations around the mean value $xn_r+x-1/2$,
which gives precise results.

\begin{figure}
\centering
\includegraphics[scale=0.5,clip]{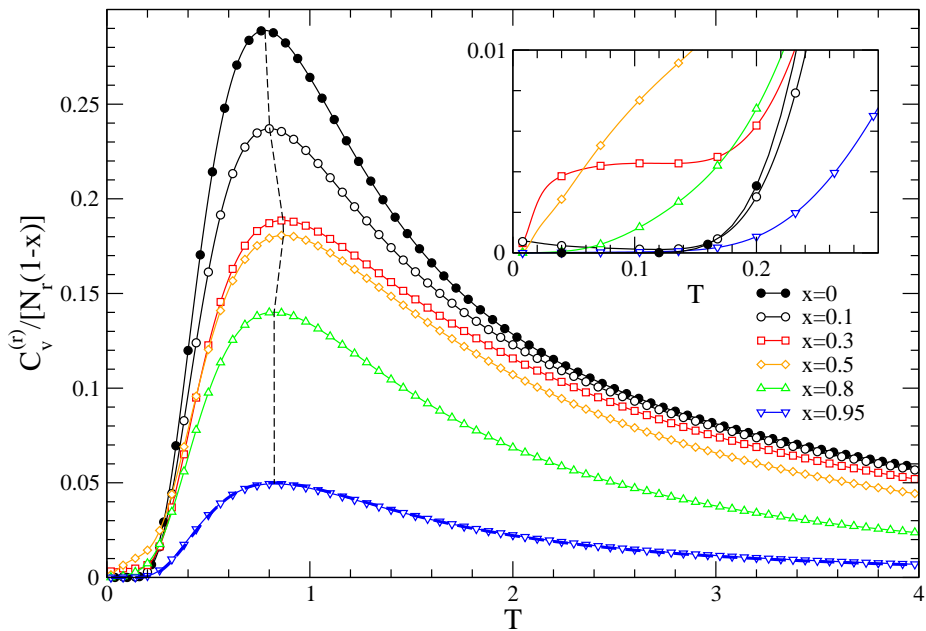}
\caption{\label{fig_cv}(Color online) Specific heat per spin $\cv{r}/[N_r(1-x)]$
at zero field
and level $r=6$ (2732 sites)
for different disorder probability values. Specific heat for the pure case $x=0$
 is derived from the exact recursion equations \eref{recurs_pure} given
in \ref{app1}. The dashed blue line at $x=0.95$ is the fit with a
two-level model which accounts for the Schottky anomaly at a temperature close
to unity, see text and \eref{cvapp}. Inset: low temperature behavior where a
plateau is visible at $x=0.3$.
The dashed black line indicates the position of each Schottky peak as function
of dilution.}
\end{figure}

\begin{figure}
\centering
\includegraphics[scale=0.5,clip]{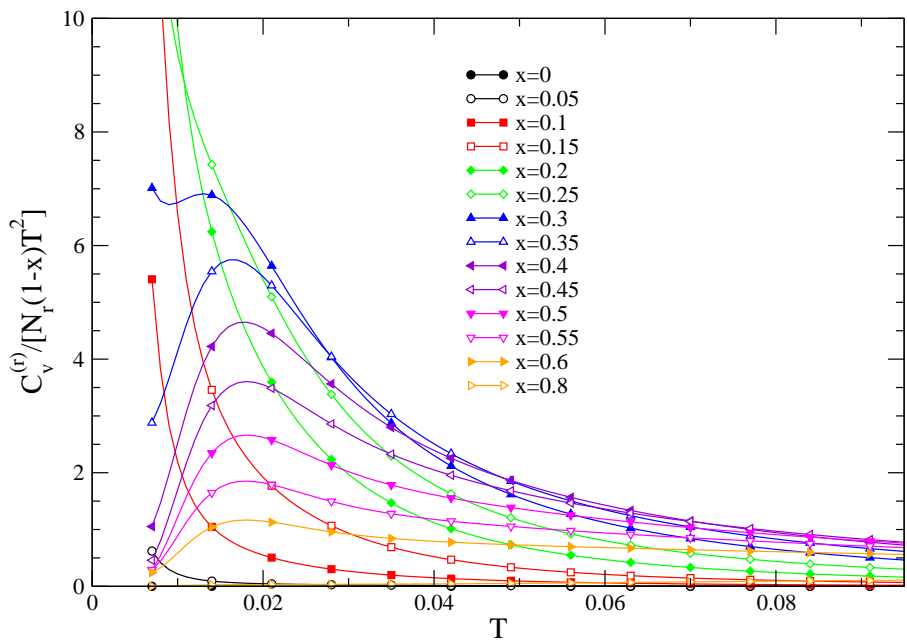}
\caption{\label{fig_cvLT}(Color online) Low temperature behavior of specific
heat per spin
$\cv{r}/[N_r(1-x)T^2]$ at zero field and level
$r=6$ (2732 sites)  for different disorder probability values. A local maximum
is developing for low dilution
$0.1\le x\le 0.3$, then non exponential behavior is observed for intermediate
values $x\simeq 0.5$.}
\end{figure}

\begin{figure}
\centering
\includegraphics[scale=0.7,angle=270,clip]{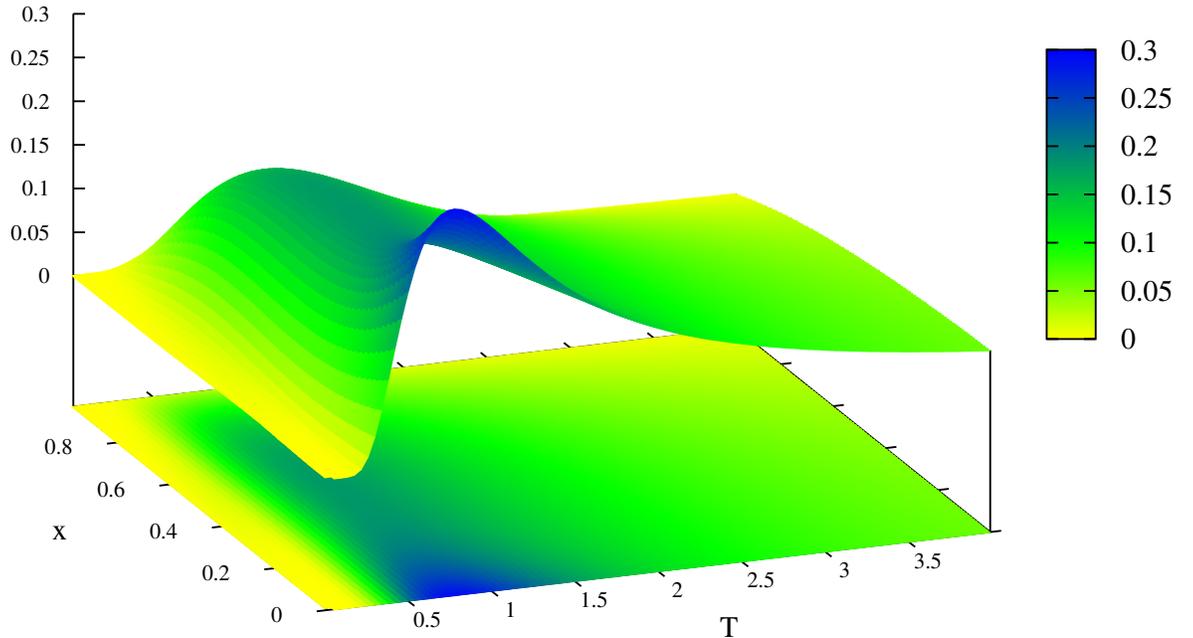}
\caption{\label{fig_cv3d}(Color online) Surface plot of the specific heat per
spin $\cv{r}/[N_r(1-x)]$ at zero field and level $r=6$ (2732 sites)
as function of temperature and dilution factor $x$. The Schottky peak amplitude
is reduced as $x$ increases.}
\end{figure}

\section{Calorimetry and thermodynamical functions in the dilute case}

In this section, we evaluate different thermodynamical quantities as function
of dilution using \eref{freeFr}.
The residual entropy per spin is plotted as function of $x$ in
\efig{fig_SvsxRes} for $r$ between 1 and 6. For the single
tetrahedron structure $r=1$, the entropy is numerically identical to the exact
expression \eref{S1} and presents non-monotonic dependence with
increasing dilution. For $r$ larger, the entropy is reduced, but saturates
rapidly after $r=5$ which corresponds to 684 sites. In the limit of extreme
dilution, the entropy per spin is simply equal to $\ln(2)$ as
expected. It is interesting to compare the resulting entropy with
the Pauling estimation $S_P$ for an infinite
number of tetrahedra treated as independent as function of dilution~\cite{ke07}

\bb\label{SP}\fl
S_P(x)=\ln(2)-3x^2(1-x)\ln(2)-2x(1-x)^2\ln(4/3)-\ff(1-x)^3\ln(8/3).
\ee

Comparison between $S_P$, the experimental data for
spin-ice Dy$_{2-x}$Y$_x$Ti$_2$O$_7$ in figure 4 of reference \cite{ke07}, and
$\ent{6}$ shows very similar values at low and moderate dilution, especially the
entropy difference in the undilute case is quite small,
$S_P(0)=\ff\ln(3/2)\simeq 0.2027$ and
$\ent{6}/N_6\simeq 0.20804$ which
is an upper bond (see also inset of \efig{fig_SvsxRes}). Exact values for the
undilute hierarchical structure can be computed up to a certain order but the
entropy shows a behavior similar to spin ice models. Approximations on
pyrochlore lattice made of Ising antiferromagnet tetrahedra give a closer
value around $0.20410$~\cite{singh12}.

The specific heat $\cv{r}$ is displayed in \efig{fig_cv} as function
of temperature for five different values of
$x$. The curves presents in general a broad maximum or Schottky anomaly
at a temperature around $T=1$ corresponding to the typical coupling $J=1$ and
associated with a crossover between a low temperature spin-ice state and
paramagnetic state. The system however stays antiferromagnetic in
the low temperature regime but is highly degenerated. The ground state energy
per spin can be computed exactly for the first terms in absence of dilution
$\en{1}/4=-1/2$, $\en{2}/12=-5/6$, and the limiting value
is estimated to be $\en{r\gg 1}/N_r\simeq -0.9$ using the recurrence equations
in \ref{app1}. The main peak location behaves non-monotonically with dilution,
as for dilute compound Dy$_{2-x}$Y$_x$Ti$_2$O$_7$ in~\cite{ke07}, which results
from the non-monotonic fraction of ground states in elementary tetrahedral
structures as seen for the entropy.
At large dilution, the specific heat can accurately be fitted with a two-level
model with gap $\Delta$ and constant $C_0$, as it can be
seen in \efig{fig_cv}

\bb\label{cvapp}
C_v^{approx}=C_0\frac{\Delta^2}{T^2}\frac{\e^{\Delta/T}}{\left
(1+\e^{\Delta/T}\right )^2}.
\ee

For example, the curve for $x=0.95$ was fitted with the previous formula using
$C_0\simeq 0.114$ and $\Delta=1.988$,
which corresponds to the specific heat for a gas of dilute pairs of spins with
$C_0\simeq 2(1-x)=0.1$ and energy coupling very close to $J=\Delta/2=1$. At
lower temperature however $T\simeq 0.1$, the specific heat presents a second
broad peak at intermediate dilution factor
$x\simeq 0.3$ (see inset of \efig{fig_cv} and \efig{fig_cvLT}) which can not be
reproduced by a two-level model.

\begin{figure}
\centering
\includegraphics[scale=0.65,angle=270,clip]{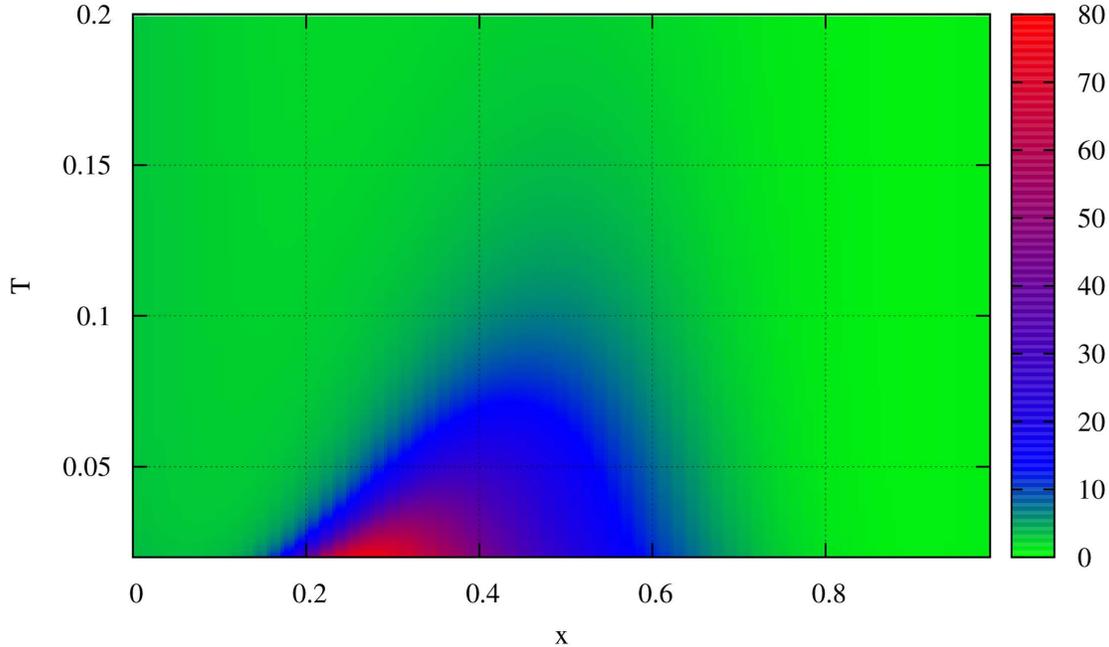}
\caption{\label{fig_delta}(Color online) Fluctuations $\DF$ of the ferromagnetic
order parameter $M$ for 2732 sites ($r=6$) as function of the dilution $x$ and
temperature.}
\end{figure}

These characteristics are exemplified in \efig{fig_cvLT} for $T\le 0.1$. The
exponential-like Arrhenius behavior of the pure system seems to evolve to more
complex features associated with a very small and broad peak contribution at
intermediate dilutions and non-exponential deviations. Arrhenius behavior is
then recovered when we approach large dilution modeled
by \eref{cvapp}. We have actually rescaled the specific heat in \efig{fig_cvLT}
by a factor $1/T^2$, in order to check if excitations like phonons
or elastic modes are present in the intermediate dilute regime.
In this case $\cv{r}$ should scale like $T^{d}$ with $d=2$ in our
model for a two-dimensional Debye contribution. Such elastic modes (in the
low temperature dynamics of domain walls for example) could result from the
non-trivial effect of long range and random distribution of the couplings, due
to the additional bonds added at each step of the lattice construction which
tend to couple remote spins and induce non-local interactions. This could
generate a random distribution of local fields, or small gaps at different
scales.

Such scaling was analyzed for example in pyrochlore compound Bi$_2$Ti$_2$O$_7$
(with $d=3$) in order to measure the excess of specific heat due to
additional Einstein oscillator contributions that could give rise to a broad
peak at low temperature~\cite{melot09}.
The scaling in $T^2$ in \efig{fig_cvLT} is more appropriate since a $T^3$
scaling would present clearly a divergence.

Non usual low-temperature specific heat behavior in dilute systems was
analyzed, in a different context, for Heisenberg magnets, within the
low-temperature spin-wave approximation~\cite{thorpe82,thorpe83}
where dilution induce non trivial temperature
exponents depending on the nature of the couplings.

To summarize, the specific heat per spin as function of both temperature $T$ and
dilution factor $x$ is displayed in \efig{fig_cv3d},
where the variation of the main Schottky peak amplitude with $x$ shows a
decreasing behavior towards a system made of individual pairs of spins with a
broader extension.

Fluctuations of the ferromagnetic order parameter defined by
$\DF=[<M^2>]_{\eta}/[(1-x)N_r]$ can be evaluated directly from the free energy
using a small field~\cite{oguchi09}

\bb\label{DF}
\DF=-\frac{T}{(1-x)N_r}\frac{\partial^2 \free{r}}{\partial h_e^2}\Big
|_{h_e=0}
+\frac{1}{(1-x)N_r}
\left (\frac{\partial \free{r}}{\partial h_e}\Big |_{h_e=0}\right )^2
\ee

and is plotted in \efig{fig_delta}. It takes noticeable values at low
temperature for intermediate dilution where short range ferromagnetic
order appears to be well developed. Such fluctuations could be associated to a
classical spin-liquid phase, as opposed to a gas state at higher temperatures
~\cite{oguchi09}.

\begin{figure}
\centering
\includegraphics[scale=0.5,angle=0,clip]{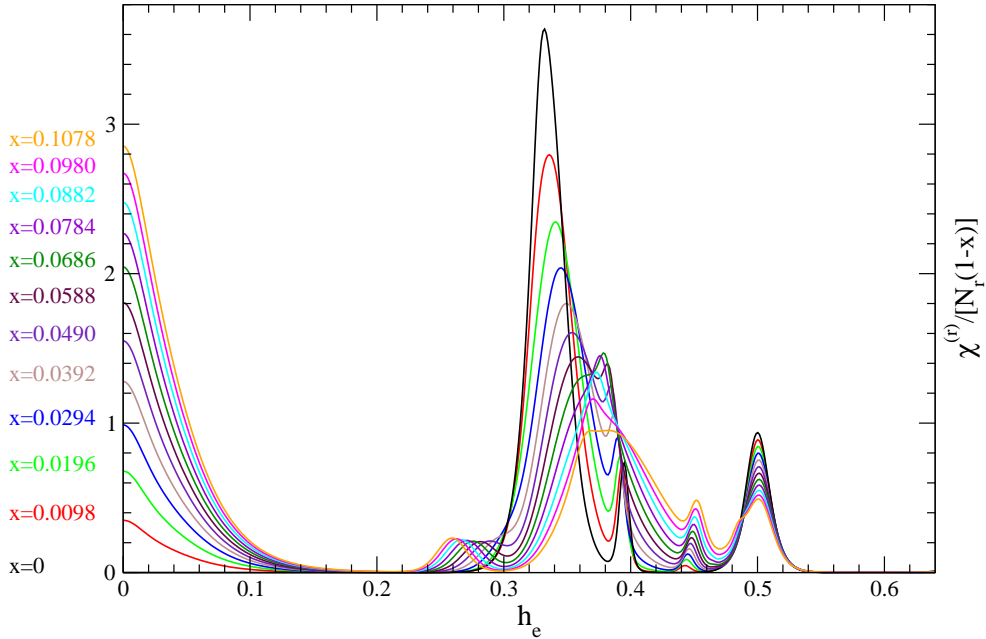}
\caption{\label{fig_chi1}(Color online) Behavior of the susceptibility
$\suscept{r}/[N_r(1-x)]$ at low temperature (T=$5.10^{-2}$) for 2732 sites
($r=6$) as function of the field $h_e$ and for several dilution factors $x$.
Here are represented only low field excitations $h_e<0.65$.
The values of $x$ are plotted on the left axis with a color corresponding to
each curve for clarity.}
\end{figure}

Susceptibility curves as function of dilution and field are plotted in
\efig{fig_chi1}. We chose to represent only the low-field excitations
$h_e<1$ in order to follow the displacement and amplitude of the first peaks
with dilution, in particular those corresponding to \efig{fig_chipure} in the
same low field region.
As dilution is increased, a new peak appears at $h_e=0$ corresponding to
excitations of uncoupled and isolated spins. The location of the peak at
$h_e=1/2$ does not change except its amplitude. It is associated to excitations
which appear numerically only at recursion level $r=4$ (172 sites), and might
probably consist in flipping two distant spins along the direction of the field,
and possibly a series of spin flippings in between, at the cost of one
frustrated link only. The energy difference between the two configurations can
be written in this case as $\Delta E=2J-4h_e$
which is negative when $h_e>1/2$. Such transition value still persists at
low dilution (less than $x=0.1$), and may result from individual un-dilute
structures with the same configurational weight, or configurations.

\begin{figure}[]
\centering
\subfigure[]
{\includegraphics[scale=0.5,angle=270,clip=true]{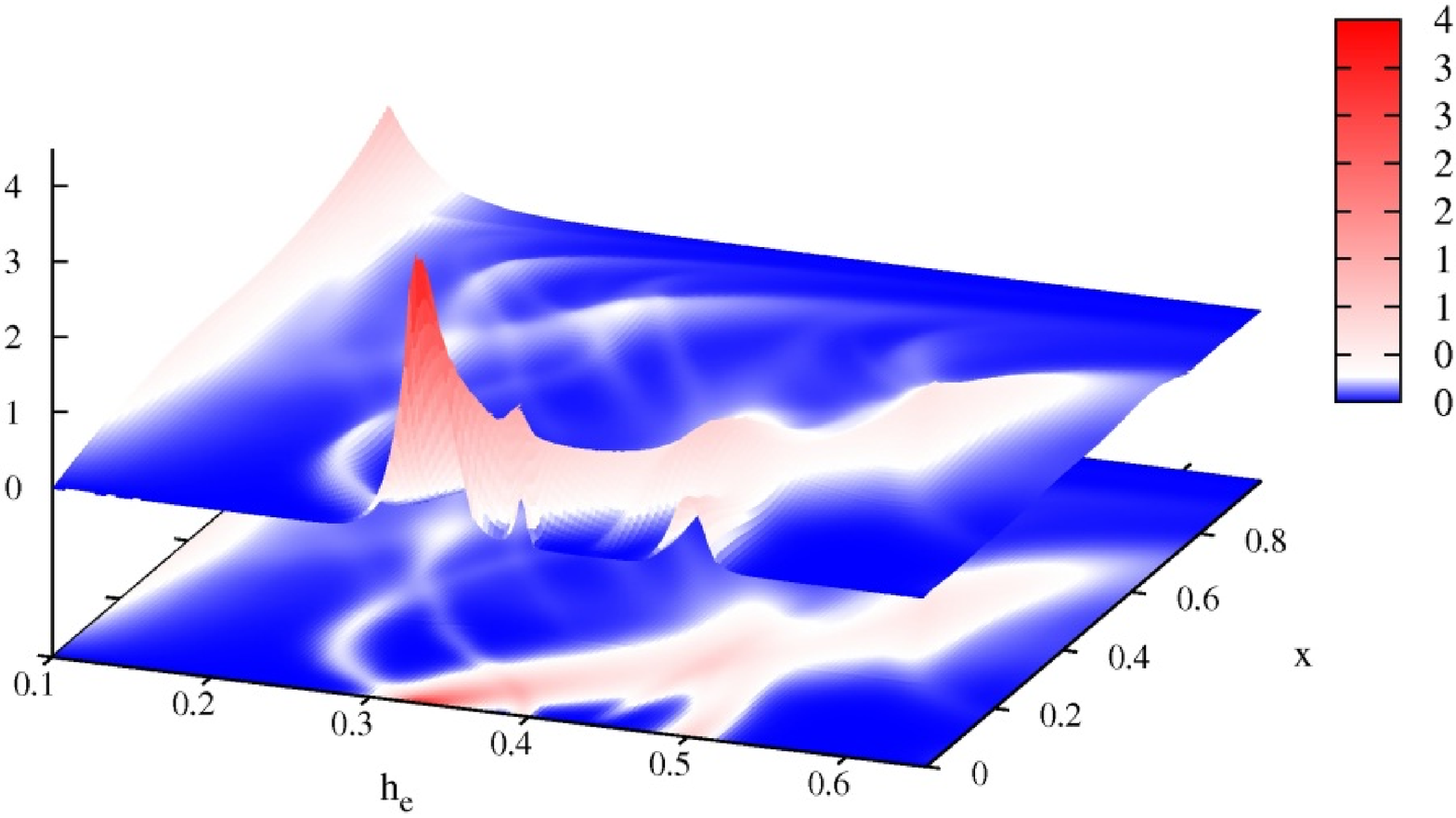}
\label{sub1}}
\subfigure[]
{\includegraphics[scale=0.465,angle=270,clip=true]{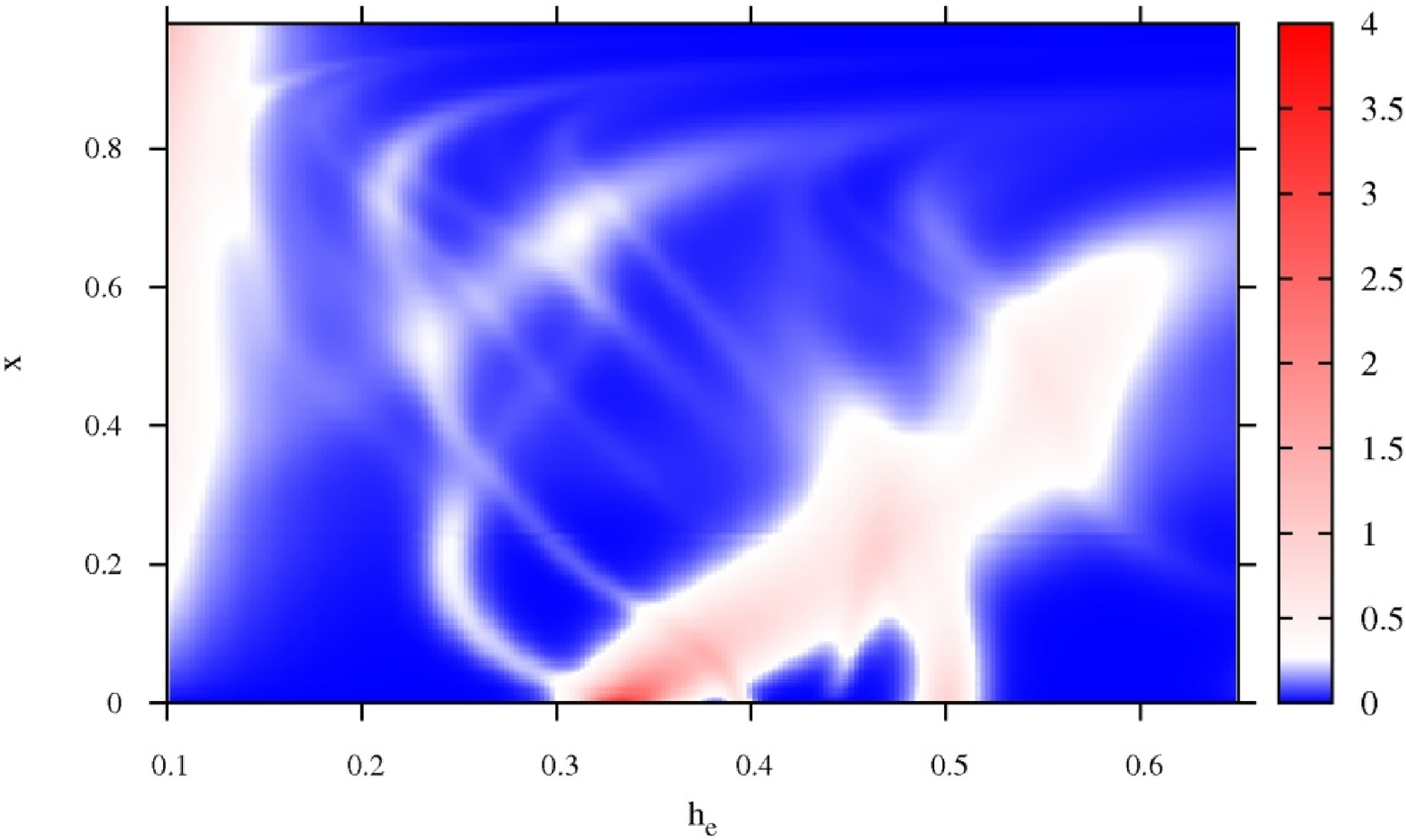}
\label{sub2}}
\caption{\label{fig_chiSurf}(Color online) (a) Surface plot representing the
spin susceptibility $\suscept{r}/[N_r(1-x)]$ at low temperature (T=$5.10^{-2}$)
for 2732 sites ($r=6$) as function of the field $h_e$ and dilution factors
$0<x<1$. Here are represented only low field excitations $0.1<h_e<0.65$
corresponding to the first three peaks of \efig{fig_chipure}. (b) Map view of
the surface plot.}
\end{figure}

The peak located at $h_e=1/3$ which appears at recursion level $r=3$ is
instead moving towards higher field values, with several intermediate peaks in
the range $1/5<h_e<1/2$.
Smaller peaks at $h_e<1/3$ are moving towards the
origin instead. A surface plot \efig{fig_chiSurf}
gives a general view of how peaks are moving with field with respect of
dilution, and how their amplitude vanishes as we approach
the high dilution regime. 
For higher field, transitions occur in small
structures of 4 spins ($r=1$) where, from a ground state of two spins up and two
spins down ($E=-2J$), transitions occur at fields $h_e=1$ and $h_e=3$
for spin flip processes corresponding successively to configurations with three
spins up, one spin down, and all spins up.

% \begin{figure}
% \includegraphics[scale=0.7,angle=270]{chiSurf.eps}
% \caption{\label{fig_chiSurf}(Color online) Surface plot representing the spin
% susceptibility $\suscept{r}/[N_r(1-x)]$ at low temperature (T=$5.10^{-2}$) for
% 2732 sites ($r=6$) as function of the field $h_e$ and dilution factors
%$0<x<1$.
% Here are represented only low field excitations $0.1<h_e<0.65$
% corresponding to the first three peaks of \efig{fig_chipure}.}
% \end{figure}

\section{Conclusion}

In this paper we propose a method to study frustrated hierarchical
lattices in presence of dilution based on the
replica method and reorganization of configurational weights at first order in
the replica parameter $n$. Interesting
properties of the dilute spin-ice state at low temperature can be examined
within this approximation by implementing recursive equations
for the partition function and leading to specific heat and susceptibility as
function of temperature and external field. Clear crossover evidence
is seen between spin-ice and paramagnetic states in the
specific heat with the presence of a Schottky peak, and the zero-temperature
entropy follows closely the Pauling approximation at least at moderate dilution.
Specific heat presents also a secondary contribution at low dilution ($x\sim
0.3$) and very low temperature with non-Arrhenius behavior, at least at the
temperatures considered numerically. This feature is
probably due to the effect of dilution on the long-range couplings across the
lattice, which involves a bimodal distribution of random antiferromagnetic
couplings between sites at different scales, and a broad distribution of random
effective fields or small gaps.

This makes this hierarchical model a good candidate for exploring
in details the physics of spin-ices or spin-liquids. Additional analysis
can probably be made using correlation functions for example or short range
ferromagnetic order parameter~\cite{oguchi09} to probe the
spin correlations in the low temperature state.
This approximation scheme based on replica may probably be implemented
more easily to hierarchical spin glass models with modal distribution of
couplings, since the quenched disorder is treated as independent between the
recursive diamond structures, making the need of a partial partition
function not necessary and therefore simplifying the analytical recurrence.
We would like to acknowledge M. Gingras for useful discussions on thermal
properties in spin-ice systems.

%---------------------------------------------------------------------------
%--------------APPENDIX-----------------------------------------------
\appendix
\section{Recursion relations for the non-disordered model\label{app1}}

In this appendix we write the recursion relations for the non-disordered case
($x=0$). Starting
from the partition function $Z_0(\sigma,\sigma')=\exp(-K\sigma\sigma')$ of a
single antiferromagnetic link between two spins $\sigma$ and $\sigma'$, and
$\ac=K$, we can generally assume the following recursive and stable form at
any step $r$

\bb\label{rec_x0}
Z_r(\sigma,\sigma')=\exp\Big(I_r-K_r\sigma\sigma'+H_r(\sigma+\sigma')\Big )
\ee

with $I_0=0$, $K_0=K$, and $H_0=0$ as initial conditions. At the next level
$r+1$, we form the
product

\bb\nn
Z_{r+1}(\sigma,\sigma')&=&\Tr{\sigma_1,\sigma_2}Z_r(\sigma,\sigma_1)Z_r(\sigma_1
,\sigma')
Z_r(\sigma,\sigma_2)Z_r(\sigma_2,\sigma')
\\
&\times&
\exp\Big (-K\sigma\sigma'-K\sigma_1\sigma_2+\he[\sigma_1+\sigma_2]\Big ).
\ee

After replacing $Z_r$ by its ansatz \eref{rec_x0}, and performing the sum over
the internal
spin degrees of freedom, we obtain

\bb\nn\fl
Z_{r+1}(\sigma,\sigma')=
\exp\Big (4I_r-K\sigma\sigma'+2H_r(\sigma+\sigma')\Big )
\left \{
2\cosh\Big [ 2K_r(\sigma+\sigma')-4H_r-2\he \Big ]\e^{-K}+2\e^{K} \right \}.
\ee

The term into bracket $\{\cdots\}$ can be rewritten as

\bb\fl
2\cosh\Big [ 2K_r(\sigma+\sigma')-4H_r-2\he \Big ]\e^{-K}+2\e^{K}=\exp
\Big (\tilde I_r-\tilde K_r\sigma\sigma'+\tilde H_r(\sigma+\sigma') \Big )
\ee

with the following equations for $\tilde I_r$, $\tilde K_r$ and $\tilde H_r$

\bb\nn
\exp(\tilde I_r-\tilde K_r+2\tilde H_r)&=&2\cosh\Big [4K_r-4H_r-2H_r \Big
]\e^{-K}+2\e^{K},
\\ \nn
\exp(\tilde I_r-\tilde K_r-2\tilde H_r)&=&2\cosh\Big [4K_r+4H_r+2H_r \Big
]\e^{-K}+2\e^{K},
\\ \nn
\exp(\tilde I_r+\tilde K_r)&=&2\cosh\Big [4K_r+2H_r \Big ]\e^{-K}+2\e^{K}.
\ee

This set of equations can be solved by eliminating successively the arguments in
the exponential terms. Then the recursive solutions
for the new couplings of $Z_{r+1}$ are given by

\bb\label{recurs_pure}
I_{r+1}=4I_r+\tilde I_r,\;K_{r+1}=K+\tilde K_r,\;H_{r+1}=2H_r+\tilde H_r.
\ee

These relations can be easily implemented in order to compute numerically the
different thermodynamical
quantities from free energy $F_r=-T\ln Z_r$.

\section{Recursion relations at finite temperature\label{app2}}

In this section the recursive equations for the different couplings in
\eref{expand} are derived. Using Kronecker integral \eref{unit},
we can rewrite $Z_{r+1}$ as

\bb\label{Zr}\fl
Z_{r+1}=\sum_{k=0}^{4n_r}
x^{k}(1-x)^{4n_r-k}
\int P(\epsilon_1)P(\epsilon_2)d\epsilon_1d\epsilon_2\Tr{\sga,\sgb}
2^{n(\epsilon_1-1)}2^{n(\epsilon_2-1)}
\\ \nn
\times \exp \left (-\ac
\epsilon\epsilon'\sum_{\alpha}\sigma^{\alpha}\sigma'^{\alpha}
-\ac \epsilon_1\epsilon_2\sum_{\alpha}\sga\sgb
-\ac
\sum_{\alpha}(\epsilon_1\sga+\epsilon_2\sgb)(\epsilon\sigma^{\alpha}
+\epsilon'\sigma'^{\alpha})
\right )
\\ \nn
\times \exp\Big [\he\sum_{\alpha}(\epsilon_1\sga+\epsilon_2\sgb)\Big]
\int_{0}^{2\pi}\frac{d\theta}{2\pi} {\rm e}^{-i\theta
k}\sum_{k_1,k_2,k_3,k_4=0}^{n_r}
\bin{n_r}{k_1}\bin{n_r}{k_2}\bin{n_r}{k_3}\bin{n_r}{k_4}
{\rm e}^{ i\theta(k_1+k_2+k_3+k_4)}
\\ \nn
\times \exp\left \{
nI_{k_1}^{(r)}(\epsilon,\epsilon_1)+nI_{k_2}^{(r)}(\epsilon_1,\epsilon')
+nI_{k_3}^{(r)}(\epsilon,\epsilon_2)+nI_{k_4}^{(r)}(\epsilon_2,\epsilon')
\right .
\\ \nn
+\left .
K_{k_1}^{(r)}\epsilon\epsilon_1
\sum_{\alpha}\sigma^{\alpha}\sigma_1^{\alpha}
+K_{k_2}^{(r)}\epsilon_1\epsilon'
\sum_{\alpha}\sigma_1^{\alpha}\sigma'^{\alpha}
+K_{k_3}^{(r)}\epsilon\epsilon_2
\sum_{\alpha}\sigma^{\alpha}\sigma_2^{\alpha}
+K_{k_4}^{(r)}\epsilon_2\epsilon'
\sum_{\alpha}\sigma_2^{\alpha}\sigma'^{\alpha}
\right .
\\ \nn
+\left .
H_{k_1}^{(r)}(\epsilon,\epsilon_1)\sum_{\alpha}(\epsilon\sigma^{\alpha}
+\epsilon_1\sigma_1^{\alpha})
+H_{k_2}^{(r)}(\epsilon',\epsilon_1)\sum_{\alpha}(\epsilon'\sigma'^{\alpha}
+\epsilon_1\sigma_1^{\alpha})
\right .
\\ \nn
+\left .
H_{k_3}^{(r)}(\epsilon,\epsilon_2)\sum_{\alpha}(\epsilon\sigma^{\alpha}
+\epsilon_2\sigma_2^{\alpha})
+H_{k_4}^{(r)}(\epsilon',\epsilon_2)\sum_{\alpha}(\epsilon'\sigma'^{\alpha}
+\epsilon_2\sigma_2^{\alpha})
\right \}.
\ee

Using \eref{identity}, we can integrate over $\theta$ and rewrite \eref{Zr} as

\bb\nn
Z_{r+1}=\sum_{k=0}^{4n_r}\bin{4n_r}{k}
x^{k}(1-x)^{4n_r-k}
\int P(\epsilon_1)P(\epsilon_2)d\epsilon_1d\epsilon_2\Tr{\sga,\sgb}
2^{n(\epsilon_1-1)}2^{n(\epsilon_2-1)}
\\ \nn
\times \exp \left (-\ac
\epsilon\epsilon'\sum_{\alpha}\sigma^{\alpha}\sigma'^{\alpha}
-\ac \epsilon_1\epsilon_2\sum_{\alpha}\sga\sgb
-\ac
\sum_{\alpha}(\epsilon_1\sga+\epsilon_2\sgb)(\epsilon\sigma^{\alpha}
+\epsilon'\sigma'^{\alpha})
\right .
\\ \nn
+\left .\sum_{k_1=\max(0,k-3n_r)}^{\min(n_r,k)}n{\cal D}^{k,k_1}_{n_r,4}
\left [
I_{k_1}^{(r)}(\epsilon,\epsilon_1)+I_{k_1}^{(r)}(\epsilon_1,\epsilon')
+I_{k_1}^{(r)}(\epsilon,\epsilon_2)+I_{k_1}^{(r)}(\epsilon_2,\epsilon')\right ]
\right .
\\ \label{Zr2}
+\left .
\sum_{k_1=\max(0,k-3n_r)}^{\min(n_r,k)}{\cal
D}^{k,k_1}_{n_r,4}K_{k_1}^{(r)}\sum_{\alpha}(\epsilon\sigma^{\alpha}
+\epsilon'\sigma'^{\alpha})
(\epsilon_1\sigma_1^{\alpha}+\epsilon_2\sigma_2^{\alpha})\right .
\\ \nn
+\left .
\sum_{k_1=\max(0,k-3n_r)}^{\min(n_r,k)}{\cal D}^{k,k_1}_{n_r,4}
\sum_{\alpha}
\left [
H_{k_1}^{(r)}(\epsilon,\epsilon_1)(\epsilon\sigma^{\alpha}+\epsilon_1\sigma_1^{
\alpha})
+H_{k_1}^{(r)}(\epsilon',\epsilon_1)(\epsilon'\sigma'^{\alpha}
+\epsilon_1\sigma_1^{\alpha})
\right . \right.
\\ \nn
+\left .\left .
H_{k_1}^{(r)}(\epsilon,\epsilon_2)(\epsilon\sigma^{\alpha}+\epsilon_2\sigma_2^{
\alpha})
+H_{k_1}^{(r)}(\epsilon',\epsilon_2)(\epsilon'\sigma'^{\alpha}
+\epsilon_2\sigma_2^{\alpha})
\right ]
+\he\sum_{\alpha}(\epsilon_1\sga+\epsilon_2\sgb) \right ].
\ee

Now the sum over intermediate spins $\sga$ and $\sgb$ can be performed directly.
Let
first define intermediate couplings

\bb
\Kr{r}{k}:=\sum_{k_1=\max(0,k-3n_r)}^{\min(n_r,k)}{\cal
D}^{k,k_1}_{n_r,4}K_{k_1}^{(r)}-\ac,
\ee

and new functions

\bb\nn
\Ir{r}{k}(\epsilon,\epsilon',\epsilon_1,\epsilon_2)&:=&\sum_{k_1=\max(0,k-3n_r)}
^{\min(n_r,k)}{\cal D}^{k,k_1}_{n_r,4}
\left [
I_{k_1}^{(r)}(\epsilon,\epsilon_1)+I_{k_1}^{(r)}(\epsilon_1,\epsilon')
+I_{k_1}^{(r)}(\epsilon,\epsilon_2)+I_{k_1}^{(r)}(\epsilon_2,\epsilon')\right ],
\\
\Hr{r}{k}(\epsilon,\epsilon',\epsilon_1)&:=&\sum_{k_1=\max(0,k-3n_r)}^{\min(n_r,
k)}{\cal D}^{k,k_1}_{n_r,4}
\left [
H_{k_1}^{(r)}(\epsilon,\epsilon_1)+H_{k_1}^{(r)}(\epsilon',\epsilon_1)
\right ].
\ee

Then we isolate the part in \eref{Zr2} containing only $\sga$ and $\sgb$, and
perform the sum:

\bb\fl
 \Tr{\sga,\sgb}
\exp \left (
-\ac \epsilon_1\epsilon_2\sum_{\alpha}\sga\sgb
+\Kr{r}{k}\sum_{\alpha}(\epsilon_1\sga+\epsilon_2\sgb)(\epsilon\sigma^{\alpha}
+\epsilon'\sigma'^{\alpha})
\right .
\\ \nn\fl
+\left .\Big [\Hr{r}{k}(\epsilon,\epsilon',\epsilon_1)+\he\Big
]\sum_{\alpha}\epsilon_1\sga
+\Big [\Hr{r}{k}(\epsilon,\epsilon',\epsilon_2)+\he\Big
]\sum_{\alpha}\epsilon_2\sgb
\right )
\\ \nn\fl
=\prod_{\alpha}\left \{
\exp(-\ac \epsilon_1\epsilon_2)2\cosh
\Big [
\Kr{r}{k}(\epsilon_1+\epsilon_2)(\epsilon\sigma^{\alpha}+\epsilon'\sigma'^{
\alpha})
+\Big (\Hr{r}{k}(\epsilon,\epsilon',\epsilon_1)+\he\Big )\epsilon_1+
\Big (\Hr{r}{k}(\epsilon,\epsilon',\epsilon_2)+\he\Big )\epsilon_2
\Big ]
\right .
\\ \nn\fl
+\left .
\exp(\ac \epsilon_1\epsilon_2)
2\cosh
\Big [
\Kr{r}{k}(\epsilon_1-\epsilon_2)(\epsilon\sigma^{\alpha}+\epsilon'\sigma'^{
\alpha})
+\Big (\Hr{r}{k}(\epsilon,\epsilon',\epsilon_1)+\he\Big )\epsilon_1-\Big (
\Hr{r}{k}(\epsilon,\epsilon',\epsilon_2)+\he\Big )\epsilon_2
\Big ]
\right \}.
\ee

Next, we perform  the integration over $\epsilon_1$ and $\epsilon_2$

\bb\nn\fl
Z_{r+1}=\sum_{k=0}^{4n_r}\bin{4n_r}{k}
x^{k}(1-x)^{4n_r-k}\exp \left (-\ac
\epsilon\epsilon'\sum_{\alpha}\sigma^{\alpha}\sigma'^{\alpha}
\right )
\\ \nn\fl
\left [
(1-x)^2\exp \left (n I_k(\epsilon,\epsilon',1,1)
+\Hr{r}{k}(1,1,1)\sum_{\alpha}(\epsilon\sigma^{\alpha}+\epsilon'\sigma'^{\alpha}
)\right )
\right .
\\ \nn\fl
\times\left .2^n
\prod_{\alpha}\left \{
\exp(-\ac )
\cosh\Big
[2\Kr{r}{k}(\epsilon\sigma^{\alpha}+\epsilon'\sigma'^{\alpha})+2\Hr{r}{k}
(\epsilon,\epsilon',1)+2\he
\Big ]+\exp(\ac )
\right \} \right .
\\ \nn\fl
+2x(1-x)\exp \left (n I_k(\epsilon,\epsilon',0,1)
+\Hr{r}{k}(1,0,1)\sum_{\alpha}(\epsilon\sigma^{\alpha}+\epsilon'\sigma'^{\alpha}
)\right )
\\ \nn\fl
\times 2^n
\prod_{\alpha}
\cosh\Big
[\Kr{r}{k}(\epsilon\sigma^{\alpha}+\epsilon'\sigma'^{\alpha})+\Hr{r}{k}(\epsilon
,\epsilon',1)+\he\Big ]
\\ \label{recurs}
+\left .x^2\exp \left (n
I_k(\epsilon,\epsilon',0,0)+\Hr{r}{k}(0,0,1)\sum_{\alpha}(\epsilon\sigma^{\alpha
}+\epsilon'\sigma'^{\alpha})\right )
\right ].
\ee

We introduce now a set of functions
$\{M_{k,l}(\epsilon,\epsilon'),Q_{k,l},H_{k,l}(\epsilon,\epsilon')\}$
for each of the three terms appearing in the previous expression and
proportional to $(1-x)^2$ ($l=0$), $2x(1-x)$ ($l=1$), and $x^2$ ($l=2$)
respectively. The first factor associated with $(1-x)^2$ can be exponentiated
such that

\bb\nn
& &\exp \left (I_k(\epsilon,\epsilon',1,1)
+\Hr{r}{k}(1,1,1)(\epsilon\sigma^{\alpha}+\epsilon'\sigma'^{\alpha})\right )
\\ \nn
&\times& 2\left \{
\exp(-\ac )
\cosh\Big
[2\Kr{r}{k}(\epsilon\sigma^{\alpha}+\epsilon'\sigma'^{\alpha})+2\Hr{r}{k}
(\epsilon,\epsilon',1)+2\he
\Big ]+\exp(\ac )
\right \}
\\
&=&:\exp \Big (
M_{k,0}(\epsilon,\epsilon')+Q_{k,0}\epsilon\epsilon'\sigma^{\alpha}\sigma'^{
\alpha}
+H_{k,0}(\epsilon,\epsilon')(\epsilon\sigma^{\alpha}+\epsilon'\sigma'^{\alpha}
)\Big ).
\ee

Similarly we have for the second term proportional to $2x(1-x)$

\bb \nn\fl
\exp \left (I_k(\epsilon,\epsilon',0,1)
+\Hr{r}{k}(1,0,1)(\epsilon\sigma^{\alpha}+\epsilon'\sigma'^{\alpha})\right )
2\cosh\Big
[\Kr{r}{k}(\epsilon\sigma^{\alpha}+\epsilon'\sigma'^{\alpha})+\Hr{r}{k}(\epsilon
,\epsilon',1)+\he\Big ]
\\ \fl
=:\exp \Big (
M_{k,1}(\epsilon,\epsilon')+Q_{k,1}\epsilon\epsilon'\sigma^{\alpha}\sigma'^{
\alpha}
+H_{k,1}(\epsilon,\epsilon')(\epsilon\sigma^{\alpha}+\epsilon'\sigma'^{\alpha}
)\Big ).
\ee

The last term associated with $x^2$ can be exponentiated using the values
$M_{k,2}(\epsilon,\epsilon'):=I_k(\epsilon,\epsilon',0,0)$, $Q_{k,2}:=0$
and $H_{k,2}(\epsilon,\epsilon'):=\Hr{r}{k}(0,0,1)$. All these functions can be
identified in a unique way by using the four possible configurations for
$\sigma^{\alpha}$ and $\sigma'^{\alpha}$. Then the partition function can be
rewritten as

\bb\fl
Z_{r+1}=\sum_{k=0}^{4n_r}\bin{4n_r}{k}
x^{k}(1-x)^{4n_r-k}\exp \left (-\ac
\epsilon\epsilon'\sum_{\alpha}\sigma^{\alpha}\sigma'^{\alpha}
\right )
\\ \nn\fl
\times\left [ \sum_{l=0}^{2}\bin{2}{l}
x^{l}(1-x)^{2-l}\exp
\Big (
nM_{k,l}(\epsilon,\epsilon')+Q_{k,l}\epsilon\epsilon'\sum_{\alpha}\sigma^{\alpha
}\sigma'^{\alpha}
+H_{k,l}(\epsilon,\epsilon')\sum_{\alpha}(\epsilon\sigma^{\alpha}
+\epsilon'\sigma'^{\alpha})\Big )
\right ].
\ee

As before we can expand the exponential terms at first order in $n$ and
rearrange the powers in $x$
such that

\bb\fl
Z_{r+1}(\epsilon\sigma^{\alpha},\epsilon'\sigma'^{\alpha})
=
\exp \left (-\ac \epsilon\epsilon'\sum_{\alpha}\sigma^{\alpha}\sigma'^{\alpha}
\right )
\sum_{k=0}^{4n_r+2=n_{r+1}}\bin{4n_r+2}{k}x^{k}(1-x)^{4n_r+2-k}\times
\\ \nn\fl
\exp
\Big (
\sum_{l=\max(0,k-4n_r)}^{\min(2,k)}\frac{\bin{4n_r}{k-l}\bin{2}{l}}{\bin{4n_r+2}
{k}}
\Big [
nM_{k-l,l}(\epsilon,\epsilon')+Q_{k-l,l}\epsilon\epsilon'\sum_{\alpha}\sigma^{
\alpha}\sigma'^{\alpha}
+H_{k-l,l}(\epsilon,\epsilon')\sum_{\alpha}(\epsilon\sigma^{\alpha}
+\epsilon'\sigma'^{\alpha})
\Big ]
\Big ).
\ee

From this result, we can deduce finally the recursive equation for the couplings

\bb\nn
K^{(r+1)}_k=\sum_{l=\max(0,k-4n_r)}^{\min(2,k)}\frac{\bin{4n_r}{k-l}\bin{2}{l}}{\bin{4n_r+2}{k}}Q_{k-l,l},
\\ \nn
I^{(r+1)}_k(\epsilon,\epsilon')=
\sum_{l=\max(0,k-4n_r)}^{\min(2,k)}\frac{\bin{4n_r}{k-l}\bin{2}{l}}{\bin{4n_r+2}{k}}M_{k-l,l}(\epsilon,\epsilon'),
\;\;{\rm and}
\\ \label{newcouplings}
H^{(r+1)}_k(\epsilon,\epsilon')=
\sum_{l=\max(0,k-4n_r)}^{\min(2,k)}\frac{\bin{4n_r}{k-l}\bin{2}{l}}{\bin{4n_r+2}{k}}H_{k-l,l}(\epsilon,\epsilon').
\ee

%-----------------bibliography---------------------------------
\section*{References}
\bibliography{biblio}

\end{document}